\documentclass[aps,preprint]{revtex4}%
\usepackage{amsfonts}
\usepackage{amsmath}
\usepackage{amssymb}
\usepackage{subfigure}
\usepackage{graphicx}
\usepackage{array}%
\begin{document}
\preprint{CTP-SCU/2018003}
\title{Thermodynamics and Phase Transitions of Nonlinear Electrodynamics Black Holes
in an Extended Phase Space}
\author{Peng Wang}
\email{pengw@scu.edu.cn}
\author{Houwen Wu}
\email{iverwu@scu.edu.cn}
\author{Haitang Yang}
\email{hyanga@scu.edu.cn}
\affiliation{Center for Theoretical Physics, College of Physical Science and Technology,
Sichuan University, Chengdu, 610064, China}

\begin{abstract}
We investigate the thermodynamic behavior of nonlinear electrodynamics (NLED)
black holes in an extended phase space, which includes the cosmological
constant $\Lambda=-3/l^{2}$ and dimensionful couplings $a$ in NLED as
thermodynamic variables. For a generic NLED black hole with the charge $Q$, we
find that the Smarr relation is satisfied in the extended phase space, and the
state of equation can be written as $Tl=\tilde{T}\left(  r_{+}/l,Q/l,al^{-c}%
\right)  $, where $\left[  a\right]  =L^{c}$, and $T$ and $r_{+}$ are the
temperature and horizon radius of the black hole, respectively. For some
values of $Q/l$ and $al^{-c}$, the phase structure of the black hole is
uniquely determined. Focusing on Born-Infeld and iBorn-Infeld AdS black holes,
we obtain the corresponding phase diagrams in the $a/l^{2}$-$Q/l$ plane, which
provides a new viewpoint towards the black holes' phase structure and critical
behavior. For Born-Infeld black holes, the critical line and the region, where
a reentrant phase transition occurs, in the $a/l^{2}$-$Q/l$ plane are both
finite and terminate at $\left\{  \tilde{a}_{c}\text{, }\tilde{Q}_{c}\right\}
\simeq\left\{  0.069\text{, }0.37\right\}  $. However for iBorn-Infeld black
holes, the critical line and the reentrant phase transition region in the
$a/l^{2}$-$Q/l$ plane are semi-infinite and extend to $Q/l=\infty$. We also
examine thermal and electrical stabilities of Born-Infeld and iBorn-Infeld AdS
black holes.

\end{abstract}
\maketitle
\tableofcontents



\section{Introduction}

Black holes are among the most intriguing concepts of general relativity,
which could have a deep impact upon the understanding of quantum gravity.
Understanding the statistical mechanics of black holes has been a subject of
intensive study for several decades. In the pioneering work
\cite{IN-Hawking:1974sw,IN-Bekenstein:1972tm,IN-Bekenstein:1973ur}, Hawking
and Bekenstein found that black holes possess the temperature and the entropy.
Analogous to the laws of thermodynamics, the four laws of black hole mechanics
were established in \cite{IN-Bardeen:1973gs}.

Studying the phase transitions of AdS black holes is primarily motivated by
AdS/CFT correspondence \cite{IN-Maldacena:1997re}. Hawking and Page showed
that a first-order phase transition occurs between Schwarzschild AdS black
holes and thermal AdS space \cite{IN-Hawking:1982dh}, which was later
understood as a confinement/deconfinement phase transition in the context of
the AdS/CFT correspondence \cite{IN-Witten:1998zw}. For Reissner-Nordstrom
(RN) AdS black holes, authors of \cite{IN-Chamblin:1999tk,IN-Chamblin:1999hg}
showed that their critical behavior is similar to that of a Van der Waals
liquid gas phase transition.

Later, the asymptotically AdS black holes have been studied in the context of
extended phase space thermodynamics, where the cosmological constant is
interpreted as thermodynamic pressure
\cite{IN-Dolan:2011xt,IN-Kubiznak:2012wp}. In this case, the black hole mass
should be understood as enthalpy instead of the internal energy
\cite{IN-Kastor:2009wy}. The $P$-$V$ criticality study has been explored for
various AdS black holes
\cite{IN-Wei:2012ui,IN-Cai:2013qga,IN-Xu:2014kwa,IN-Frassino:2014pha,IN-Dehghani:2014caa,IN-Hennigar:2015esa}%
. It showed that the $P$-$V$ critical behaviors of AdS black holes are similar
to that of a Van der Waals liquid gas system. A reentrant phase transition
occurs if, as one monotonically changes a thermodynamic variable, the system
undergoes two (or more) phase transitions and returns to a state
macroscopically similar to the initial state. In the context of the extended
phase space, the reentrant phase transition has been observed for some AdS
black holes, e.g., 4D BI-AdS black holes \cite{IN-Gunasekaran:2012dq}, higher
dimensional singly spinning Kerr-AdS black holes \cite{IN-Altamirano:2013ane},
AdS black holes in Lovelock gravity \cite{IN-Frassino:2014pha}, AdS black
holes in dRGT massive gravity \cite{IN-Zou:2016sab}, hairy AdS black holes
\cite{IN-Hennigar:2015wxa}.\

Nonlinear electrodynamics (NLED) is an effective model incorporating quantum
corrections to Maxwell electromagnetic theory. Coupling NLED to gravity,
various NLED charged black holes were derived and discussed in a number of
papers
\cite{IN-Soleng:1995kn,IN-AyonBeato:1998ub,IN-Maeda:2008ha,IN-Hendi:2017mgb,IN-Tao:2017fsy,IN-Guo:2017bru,IN-Mu:2017usw}%
. The thermodynamics of NLED black holes in the extended phase space have been
considered in literature, e.g., power Maxwell invariant black holes
\cite{IN-Hendi:2012um,IN-Mo:2016jqd,IN-Yerra:2018mni}, non-linear
magnetic-charged dS black hole \cite{IN-Nam:2018tpf}.

Among the various NLED, there is a famous string-inspired one: Born-Infeld
electrodynamics, which encodes the low-energy dynamics of D-branes.
Born-Infeld electrodynamics incorporates maximal electric fields and smooths
divergences of the electrostatic self-energy of point charges. The Born-Infeld
AdS (BI-AdS) black hole solution was first obtained in
\cite{IN-Dey:2004yt,IN-Cai:2004eh}. The thermodynamic behavior and phase
transitions of BI-AdS black holes were studied in the canonical ensemble
\cite{IN-Fernando:2003tz} and in the grand ensemble \cite{IN-Fernando:2006gh}.
The critical behavior and thermodynamics of BI-AdS black holes in various
gravities were also investigated in
\cite{IN-Banerjee:2010da,IN-Banerjee:2011cz,IN-Lala:2011np,IN-Banerjee:2012zm,IN-Azreg-Ainou:2014twa,IN-Hendi:2015hoa,IN-Zangeneh:2016fhy,IN-Zeng:2016sei,IN-Li:2016nll}%
. In the extended phase space, the thermodynamic phase structure and critical
behavior of 4D and higher dimensional BI-AdS black holes were studied in
\cite{IN-Gunasekaran:2012dq,IN-Zou:2013owa}, respectively. In
\cite{IN-Dehyadegari:2017hvd}, the thermodynamics of 4D BI-AdS black holes has
recently been discussed in the case, in which the charge of the black hole
varies and the pressure is fixed. The reentrant phase transition has been
observed in 4D BI-AdS black holes while there was no reentrant phase
transition for the system of higher dimensional BI-AdS black holes. Although
the properties of BI-AdS black holes are thoroughly investigated in
literature, their electrical stabilities have been rarely reported.

Recently, a new type of NLED black holes, namely iBorn-Infeld AdS (iBI-AdS)
black holes, have been considered in \cite{IN-Baggioli:2016oju} as holographic
models behaving as prototypes of Mott insulators. The Lagrangian of the
iBorn-Infeld field can be obtained from that of the Born-Infeld field by
extending the BI parameter ($a$ in eqn. $\left(  \ref{eq:BI}\right)  $) to a
negative real number. In \cite{IN-Wang:2018hwg}, it showed that the
nonlinearity correction tends to reduce/increase the strength of the repulsive
force between two electrons for the Born-Infeld/iBorn-Infeld field. So it is
natural to expect that the iBI-AdS black hole is dual to a theory with strong
interactions between electrons, which could lead to Mott-like behavior.
Moreover, negative magneto-resistance and the Mott insulator to metal
transition induced by a magnetic field can be realized at low temperatures in
the iBorn-Infeld holographic models. Compared to the Born-Infeld case, the
iBorn-Infeld case leads to a much richer transport behavior in the dual
theory. As shown in \cite{IN-Baggioli:2016oju}, iBI-AdS black holes satisfy
the constraints to ensure consistency in the form of ghosty perturbations
and/or gradient instabilities at the decoupling limit. However, the
thermodynamic behavior and phase structure of iBI-AdS black holes have yet to
be discussed.

In this paper, we first investigate the thermodynamic behavior of\ generic
NLED black holes in the extended phase space. Then, we turn to study the phase
structure and critical behavior of BI-AdS and iBI-AdS black holes by studying
the phase diagrams in the $Q/l$-$a/l^{2}$ plane. The rest of this paper is
organized as follows. In section \ref{Sec:NBH}, we derive the NLED black hole
solution, compute its Euclidean action and discuss thermodynamic properties of
the black hole. We find that the Smarr relation is satisfied after including
dimensionful couplings in NLED in the extended phase space. In section
\ref{Sec:BIABH}, we study the phase structure and critical behavior of BI-AdS
black holes. The phase diagram for BI-AdS black holes in the $Q/l$-$a/l^{2}$
plane is given in FIG. \ref{fig:DBIQa}, from which one can read the black
hole's phase structure and critical behavior. We further explore thermal and
electrical stabilities of BI-AdS black holes. In section \ref{Sec:iBIABH}, the
phase structure and critical behavior of iBI-AdS black holes are investigated,
which can be inferred from the phase diagram in the $Q/l$-$a/l^{2}$ plane,
FIG. \ref{fig:iDBIQa}. We also study thermal and electrical stabilities of
iBI-AdS black holes. We summarize our results in section \ref{Sec:Con}. In
appendix, we present an alternative derivation of the Smarr relation for NLED
black holes.

\section{NLED Black Hole}

\label{Sec:NBH}

In this section, we first derive the asymptotically AdS black hole solution in
the Einstein-NLED gravity. After its Gibbs free energy is obtained via
calculating the Euclidean action, we then discuss the thermodynamic properties
of the black hole, e.g., Smarr relation, stability.

\subsection{Black Hole Solution}

Consider a 4-dimensional model of gravity coupled to a nonlinear
electromagnetic field $A_{\mu}$ with the action given by%
\begin{equation}
S_{\text{Bulk}}=\int d^{4}x\sqrt{-g}\left[  R-2\Lambda+\mathcal{L}\left(
s,a_{i}\right)  \right]  , \label{eq:Action}%
\end{equation}
where the cosmological constant $\Lambda=-\frac{3}{l^{2}}$, and we take $16\pi
G=1$ for simplicity. In the action $\left(  \ref{eq:Action}\right)  $, we
assume that the generic NLED Lagrangian $\mathcal{L}\left(  s,a_{i}\right)  $
is a function of $s$ and the parameters $a_{i}$, where we build an independent
nontrivial scalar using $F_{\mu\nu}=\partial_{\mu}A_{\nu}-\partial_{\nu}%
A_{\mu}$ and none of its derivatives:
\begin{equation}
s=-\frac{1}{4}F^{\mu\nu}F_{\mu\nu}\text{.}%
\end{equation}
The parameters $a_{i}$ characterize the effects of nonlinearity in the NLED.
We also assume that the NLED Lagrangian would reduce to the Maxwell Lagrangian
for small fields:%
\begin{equation}
\mathcal{L}\left(  s,a_{i}\right)  \approx s.
\end{equation}
Varying the action $\left(  \ref{eq:Action}\right)  $ with respect to $g_{ab}$
and $A_{a}$, we find that the equations of motion are%
\begin{align}
R_{\mu\nu}-\frac{1}{2}Rg_{\mu\nu}-\frac{3}{l^{2}}g_{\mu\nu}  &  =\frac
{T_{\mu\nu}}{2}\text{,}\nonumber\\
\nabla_{\mu}G^{\mu\nu}  &  =0\text{,}%
\end{align}
where $T_{\mu\nu}$ is the energy-momentum tensor:%
\begin{equation}
T_{\mu\nu}=g_{\mu\nu}\mathcal{L}\left(  s,a_{i}\right)  +\frac{\partial
\mathcal{L}\left(  s,a_{i}\right)  }{\partial s}F_{\mu}^{\text{ }\rho}%
F_{\nu\rho}\text{,}%
\end{equation}
and we define the auxiliary fields $G^{\mu\nu}$:
\begin{equation}
G^{\mu\nu}=-\frac{\partial\mathcal{L}\left(  s,a_{i}\right)  }{\partial
F_{\mu\nu}}=\frac{\partial\mathcal{L}\left(  s,a_{i}\right)  }{\partial
s}F^{\mu\nu}. \label{eq:Gab}%
\end{equation}
\

To construct a black hole solution with asymptotic AdS spacetime, we take the
following ansatz for the metric and the NLED field%
\begin{align}
ds^{2}  &  =-f\left(  r\right)  dt^{2}+\frac{dr^{2}}{f\left(  r\right)
}+r^{2}\left(  d\theta^{2}+\sin^{2}\theta d\phi^{2}\right)  \text{,}%
\nonumber\\
A  &  =A_{t}\left(  r\right)  dt\text{.} \label{eq:ansatz}%
\end{align}
The equations of motion then take the form:%
\begin{align}
-1+f\left(  r\right)  -\frac{3r^{2}}{l^{2}}+rf^{\prime}\left(  r\right)   &
=\frac{r^{2}}{2}\left[  \mathcal{L}\left(  s,a_{i}\right)  +A_{t}^{\prime
}\left(  r\right)  G^{rt}\right]  ,\label{eq:ttEOM}\\
2f^{\prime}\left(  r\right)  -\frac{6r}{l^{2}}+rf^{\prime\prime}\left(
r\right)   &  =r\mathcal{L}\left(  s,a_{i}\right)  ,\label{eq:rrEOM}\\
\left[  r^{2}G^{rt}\right]  ^{\prime}  &  =0\text{,} \label{eq:NLEDEOM}%
\end{align}
where%
\begin{equation}
s=\frac{A_{t}^{\prime2}\left(  r\right)  }{2}\text{ and }G^{rt}=-\frac
{\partial\mathcal{L}\left(  s,a_{i}\right)  }{\partial s}A_{t}^{\prime}\left(
r\right)  \text{.} \label{eq:sGrt}%
\end{equation}
It can show that eqn. $\left(  \ref{eq:rrEOM}\right)  $ can be derived from
eqns. $\left(  \ref{eq:ttEOM}\right)  $ and $\left(  \ref{eq:NLEDEOM}\right)
$. Eqn. $\left(  \ref{eq:NLEDEOM}\right)  $ leads to%
\begin{equation}
G^{tr}=\frac{q}{r^{2}}\text{,} \label{eq:Grt}%
\end{equation}
where $q$ is a constant. Via eqns. $\left(  \ref{eq:sGrt}\right)  $ and
$\left(  \ref{eq:Grt}\right)  $, $A_{t}^{\prime}\left(  r\right)  $ is
determined by%
\begin{equation}
\mathcal{L}^{\prime}\left(  \frac{A_{t}^{\prime2}\left(  r\right)  }{2}%
,a_{i}\right)  A_{t}^{\prime}\left(  r\right)  =\frac{q}{r^{2}}.
\label{eq:QAt}%
\end{equation}
Moreover, integrating eqn. $\left(  \ref{eq:rrEOM}\right)  $ leads to
\begin{equation}
f\left(  r\right)  =1-\frac{m}{r}+\frac{r^{2}}{l^{2}}-\frac{1}{2r}\int
_{r}^{\infty}drr^{2}\left[  \mathcal{L}\left(  \frac{A_{t}^{\prime2}\left(
r\right)  }{2},a_{i}\right)  -A_{t}^{\prime}\left(  r\right)  \frac{q}{r^{2}%
}\right]  ,
\end{equation}
where $m$ is a constant. For large values of $r$, one finds that%
\begin{equation}
f\left(  r\right)  =1-\frac{m}{r}+\frac{r^{2}}{l^{2}}+\frac{q^{2}}{4r^{2}%
}+\mathcal{O}\left(  r^{-4}\right)  , \label{eq:f(r)LV}%
\end{equation}
which reduces to the behavior of a RN-AdS black hole. At the horizon $r=r_{+}$
where $f\left(  r_{+}\right)  =0$, the Hawking temperature of the black brane
is given by%
\begin{equation}
T=\frac{f^{\prime}\left(  r_{+}\right)  }{4\pi}\text{.}%
\end{equation}
Hence at $r=r_{+}$, eqn. $\left(  \ref{eq:ttEOM}\right)  $ gives%
\begin{equation}
T=\frac{1}{4\pi r_{+}}\left\{  1+\frac{3r_{+}^{2}}{l^{2}}+\frac{r_{+}^{2}}%
{2}\left[  \mathcal{L}\left(  \frac{A_{t}^{\prime2}\left(  r_{+}\right)  }%
{2},a_{i}\right)  -A_{t}^{\prime}\left(  r_{+}\right)  \frac{q}{r_{+}^{2}%
}\right]  \right\}  . \label{eq:HT}%
\end{equation}

The charge $Q$ of the black hole can be expressed in terms of the constants
$q$. In fact, if we turn on the the external current $J^{\mu}$, the action
would include an interaction term:
\begin{equation}
S_{I}=4\pi\int d^{4}x\sqrt{-g}J^{\mu}A_{\mu},
\end{equation}
and hence the equation of motion for $A_{\mu}$ becomes%
\begin{equation}
\nabla_{\nu}G^{\mu\nu}=4\pi J^{\mu}. \label{eq:GJ}%
\end{equation}
The charge passing through a spacelike hypersurface $\Sigma$ is given by%
\begin{equation}
Q=-\int_{\Sigma}d^{3}x\sqrt{\gamma}\sigma_{\mu}J^{\mu},
\end{equation}
where $\gamma_{ij}$ is the induced metric, and $\sigma^{\mu}$ is the unit
normal vector of $\Sigma$. Using Stokes's theorem and eqn. $\left(
\ref{eq:GJ}\right)  $, we can express the charge as a boundary integral:%
\begin{equation}
Q=-\frac{1}{4\pi}\int_{\partial\Sigma}d^{2}x\sqrt{\tilde{\gamma}}n_{\mu}%
\sigma_{\nu}G^{\mu\nu}, \label{eq:QI}%
\end{equation}
where $\partial\Sigma$ is the boundary of $\Sigma$, $\tilde{\gamma}_{ij}$ is
the induced metric, and $n_{\mu}$ is the unit outward-pointing normal vector.
For the metric in eqn. $\left(  \ref{eq:ansatz}\right)  $, $\Sigma$ and
$\partial\Sigma$ can be a constant-$t$ hypersurface and a two-sphere at
$r=\infty$, respectively. In this case, one has%
\begin{equation}
\sigma_{\mu}=\left(  -f^{1/2},0,0,0\right)  \text{ and }n_{\mu}=\left(
0,f^{-1/2},0,0\right)  \text{.}%
\end{equation}
Thus, the charge of the black hole given by eqn. $\left(  \ref{eq:QI}\right)
$ becomes%
\begin{equation}
Q=\frac{1}{4\pi}\int d\theta d\phi r^{2}\sin\theta\frac{q}{r^{2}}=q,
\end{equation}
where we use eqn. $\left(  \ref{eq:Grt}\right)  $. The gauge potential
measured with respect to the horizon is%
\begin{equation}
\Phi=4\pi\int_{r_{+}}^{\infty}A_{t}^{\prime}\left(  r\right)  =4\pi
A_{t}\left(  \infty\right)  , \label{eq:potential}%
\end{equation}
where we fix the gauge field $A_{t}\left(  r\right)  $ at the horizon to be
zero, i.e., $A_{t}\left(  r_{+}\right)  =0$. The electrostatic potential
$\Phi$ plays a role as the conjugated variable to $Q$ in black hole thermodynamics.

For asymptotically AdS spaces, the mass may be extracted by comparison to a
reference background, e.g., vacuum AdS. Similar to the charge of the black
hole, the mass can also be determined by the Komar integral%
\begin{equation}
M=4\int d\theta d\phi r^{2}\sin\theta\left(  \sigma_{\mu}n_{\nu}\nabla^{\mu
}K^{\nu}\right)  -M_{\text{AdS}},
\end{equation}
where $K^{\mu}=\left(  1,0,0,0\right)  $ is the Killing vector associated with
$t$, and $M_{\text{AdS}}$ is Komar integral associated with $K^{\mu}$ for
vacuum AdS space%
\begin{equation}
M_{\text{AdS}}=4\int d\theta d\phi r^{2}\sin\theta\left(  \frac{r}{l^{2}%
}\right)  .
\end{equation}
At spatial infinity, one can use eqn. $\left(  \ref{eq:f(r)LV}\right)  $ to
calculate%
\begin{equation}
\sigma_{\mu}n_{\nu}\nabla^{\mu}K^{\nu}=\frac{1}{2}f^{\prime}\left(  r\right)
=\frac{m}{2r^{2}}+\frac{r}{l^{2}}+\mathcal{O}\left(  r^{-3}\right)  .
\end{equation}
So the mass of the black hole is%
\begin{equation}
M=8\pi m.
\end{equation}

\subsection{Euclidean Action Calculation}

In the Euclidean path integral approach to quantum gravity
\cite{NELDBH-Hawking:1978jz,NELDBH-Gibbons:2002du}, one can identify the
Euclidean path integral with the thermal partition function:%
\[
Z=\int\mathcal{D}ge^{-S^{E}\left(  g\right)  }.
\]
In the semiclassical approximation, the dominant contribution to the path
integral comes from the classical solution, and hence one has%
\begin{equation}
Z\simeq e^{-S^{E}}.
\end{equation}
Here, $S^{E}$ is the on-shell action which is obtained by substituting the
classical solution to the action. In asymptotically AdS spaces, $S^{E}$ needs
to be regulated to cancel the divergences coming from the asymptotic region.
In the background-substraction method, one can regularize $S^{E}$ by
subtracting a contribution from a reference background. In
\cite{NELDBH-Chamblin:1999hg}, the background-substraction method was used to
compute $S^{E}$ for RN-AdS black holes. On the other hand, there is the
counterterm subtraction method
\cite{NELDBH-Balasubramanian:1999re,NELDBH-Emparan:1999pm}, in which the
action $S^{E}$ is regularized in a background-independent fashion by adding a
series of boundary counterterms to the action. Specifically, the Kounterterms
method \cite{NELDBH-Olea:2005gb,NELDBH-Olea:2006vd} has been proposed as a
regularization scheme for gravity in asymptotically AdS spaces. In
\cite{NELDBH-Miskovic:2008ck}, the Euclidean action was computed for black
hole solutions of AdS gravity coupled to the Born-Infeld electrodynamics using
the Kounterterms method.

We now follow the method in \cite{NELDBH-Miskovic:2008ck} to calculate the
Euclidean action for the asymptotically AdS NLED black hole solution $\left(
\ref{eq:ansatz}\right)  $ in canonical ensemble, in which the temperature and
charge of the black hole are fixed. The regularized action is then given by%
\begin{equation}
S_{R}=S_{\text{Bulk}}+S_{\text{ct}}+S_{\text{surf}}%
\end{equation}
where the boundary terms are%
\begin{equation}
S_{\text{ct}}=\frac{l^{2}}{4}\int d^{3}yB_{3}\text{ and }S_{\text{surf}}=-\int
d^{3}y\sqrt{\gamma}n_{\nu}G^{\mu\nu}A_{\mu},
\end{equation}
$B_{3}$ is the 2nd Chern form, and $n^{\mu}$ is the unit outward-pointing
normal vector of the boundary. Since the asymptotically AdS spacetime has
constant curvature in the asymptotic region, it showed in
\cite{NELDBH-Miskovic:2008ck} that after including the boundary term
$S_{\text{ct}}$, the action was stationary around the classical solution under
arbitrary variations of the metric $g_{\mu\nu}$. To keep the charge of the
black hole fixed instead of the potential, the boundary term $S_{\text{surf}}$
has to be added. In fact, varying the action with respect to $A_{\mu}$ gives%
\[
\delta S_{R}=\text{EOM}-\int d^{3}y\sqrt{\gamma}n_{\nu}\delta G^{\mu\nu}%
A_{\mu}=\text{EOM}-\int d^{3}y\frac{\sqrt{h}n_{r}A_{t}}{r^{2}}\delta Q.
\]

For the Euclidean continuation of the action $S^{E}=iS_{R}$, the horizon at
$r=r_{+}$ is shrunk to a point, and the manifold spans between $r=r_{+}$ and
$r=\infty$. To avoid a conical singularity at the origin of the radial
coordinate, one requires to identify the Euclidean time $\tau=-it$ as
$\tau\sim\tau+\beta$, where the period $\beta=T^{-1}$ is the inverse of the
Hawking temperature $T$. The Euclidean continuation of the counter term
$S_{\text{ct}}$ was calculated in \cite{NELDBH-Miskovic:2008ck}:%
\begin{equation}
S_{\text{ct$\;$}}^{E}=-4\pi\beta l^{2}f^{\prime}\left(  r\right)  \left[
f\left(  r\right)  -1\right]  |_{r=\infty}.
\end{equation}
The bulk action is
\begin{equation}
S_{\text{Bulk}}^{E}=-\int_{0}^{\beta}d\tau\int d\Omega\int_{r_{+}}^{\infty
}drr^{2}\left[  R+\frac{6}{l^{2}}+\mathcal{L}\left(  s,a_{i}\right)  \right]
=4\pi\beta\left[  r^{2}f^{\prime}\left(  r\right)  \right]  |_{r_{+}}^{\infty
}-\beta q\Phi,
\end{equation}
where we use eqns. $\left(  \ref{eq:rrEOM}\right)  $ and $\left(
\ref{eq:potential}\right)  $. The boundary term $S_{\text{surf}}^{E}$ is
\begin{equation}
S_{\text{surf}}^{E}=\int_{0}^{\beta}d\tau\int d\Omega f^{1/2}\left(  r\right)
r^{2}\sin\theta\left(  n_{\nu}G^{\mu\nu}A_{\mu}\right)  |_{r=\infty}=\beta
q\Phi,
\end{equation}
where we use eqn. $\left(  \ref{eq:potential}\right)  $, $f\left(  r\right)
\rightarrow1$ as $r\rightarrow\infty$, and $n_{\mu}=\left(  0,f^{-1/2}%
,0,0\right)  .$ To sum up all terms, the Euclidean action $S^{E}$ is given by
\begin{equation}
S^{E}=-16\pi^{2}r_{+}^{2}+4\pi\beta\left\{  r^{2}f^{\prime}\left(  r\right)
-l^{2}f^{\prime}\left(  r\right)  \left[  f\left(  r\right)  -1\right]
\right\}  |_{r=\infty}=\beta\left(  M-TS\right)  ,
\end{equation}
where the entropy of the black hole is
\begin{equation}
S=16\pi^{2}r_{+}^{2},
\end{equation}
and eqn. $\left(  \ref{eq:f(r)LV}\right)  $ gives%
\begin{equation}
\lim_{r\rightarrow\infty}\left\{  r^{2}f^{\prime}\left(  r\right)
-l^{2}f^{\prime}\left(  r\right)  \left[  f\left(  r\right)  -1\right]
\right\}  =2m.
\end{equation}
Since the Euclidean action is calculated at fixed $Q$, $P\left(
=6/l^{2}\right)  $ and $T$, we can associate it with the Gibbs free energy:%
\begin{equation}
F=M-TS.
\end{equation}

\subsection{Thermodynamic}

Here, we study the thermodynamics of the NLED AdS black hole solution in the
extended phase space. In such perspective on black hole thermodynamics, one
needs to include the cosmological constant $\Lambda$ as a pressure term and
interpret the mass of the black hole as a gravitational version of chemical
enthalpy. Furthermore, as noted in Lovelock gravity
\cite{NELDBH-Kastor:2010gq} and Born-Infeld electrodynamics
\cite{IN-Gunasekaran:2012dq}, any dimensionful coupling should be promoted to
a thermodynamic variable and hence introduce the associated conjugate, which
would add an extra term in the first law and Smarr relation.

In terms of the horizon radius $r_{+}$, the mass $M$ can be written as%
\begin{equation}
M=8\pi\left\{  r_{+}+\frac{r_{+}^{3}}{l^{2}}-\frac{1}{2}\int_{r_{+}}^{\infty
}drr^{2}\left[  \mathcal{L}\left(  \frac{A_{t}^{\prime2}\left(  r\right)  }%
{2},a_{i}\right)  -A_{t}^{\prime}\left(  r\right)  \frac{Q}{r^{2}}\right]
\right\}  . \label{eq:Mass}%
\end{equation}
So the derivatives of the mass in terms of the entropy and the charge are,
respectively,
\begin{equation}
\frac{\partial M}{\partial S}=\frac{1}{4\pi r_{+}}\frac{\partial m}{\partial
r_{+}}=T\text{,}%
\end{equation}
and
\begin{equation}
\frac{\partial M}{\partial Q}=8\pi\left[  -\frac{1}{2}\int_{r_{+}}^{\infty
}drr^{2}\mathcal{L}^{\prime}\left(  \frac{A_{t}^{\prime2}\left(  r\right)
}{2},a_{i}\right)  A_{t}^{\prime}\left(  r\right)  \frac{\partial
A_{t}^{\prime}\left(  r\right)  }{\partial Q}+\frac{\Phi}{8\pi}+\frac{Q}{8\pi
}\frac{\partial\Phi}{\partial Q}\right]  =\Phi,
\end{equation}
where we use eqns. $\left(  \ref{eq:QAt}\right)  $ and $\left(
\ref{eq:potential}\right)  $. Since the pressure $P=6/l^{2}$, one has%
\begin{equation}
\frac{\partial M}{\partial P}=\frac{4\pi}{3}r_{+}^{3}\equiv V\text{,}%
\end{equation}
where $V$ is the thermodynamic volume. For dimensionful couplings $a_{i}$ in
$\mathcal{L}\left(  s,a_{i}\right)  $, we can introduce the associated
conjugates $\mathcal{A}_{i}$:%
\begin{equation}
\mathcal{A}_{i}=\frac{\partial M}{\partial a_{i}}.
\end{equation}
Therefore, the extended first law takes the form%
\begin{equation}
dM=TdS+VdP+\Phi dQ+\sum\limits_{i}\mathcal{A}_{i}da_{i}. \label{eq:firstLaw}%
\end{equation}
Performing the dimensional analysis, we assume that $\left[  a_{i}\right]
=L^{c_{i}}$. The Euler scaling argument \cite{NELDBH-Kastor:2009wy} gives the
Smarr relation for the black holes%
\begin{equation}
M=2\left(  TS-VP\right)  +\sum\limits_{i}c_{i}a_{i}\mathcal{A}_{i}+Q\Phi.
\end{equation}
As a check, the Smarr relation are derived directly from the definitions of
the thermodynamic quantities of the black hole in the appendix.

Till now, our expressions for the thermodynamics quantities, e.g., the Gibbs
free energy $F$, the enthalpy $M$, are functions of the horizon radius $r_{+}$
(the entropy $S$), the charge $Q$ and the pressure $P$ (the AdS radius $l$).
However, in canonical ensemble with fixed $T$, $Q$ and $P$, we need to express
the thermodynamics quantities in terms of $T$, $Q$ and $P$. In doing so, the
equation of state $\left(  \ref{eq:HT}\right)  $ is solved for $r_{+}$:
$r_{+}=r_{+}\left(  T,Q,P,a_{i}\right)  $. Interestingly, the equation of
state $\left(  \ref{eq:HT}\right)  $ can be rewritten as%
\begin{equation}
\tilde{T}=\frac{1}{4\pi\tilde{r}_{+}}\left\{  1+3\tilde{r}_{+}^{2}+\frac{1}%
{2}\tilde{r}_{+}^{2}\left[  \mathcal{L}\left(  \frac{\tilde{A}_{t}^{\prime
2}\left(  \tilde{r}_{+}\right)  }{2},\tilde{a}_{i}\right)  -\tilde{A}%
_{t}^{\prime}\left(  \tilde{r}_{+}\right)  \frac{\tilde{Q}}{\tilde{r}_{+}^{2}%
}\right]  \right\}  , \label{eq:Ttildal}%
\end{equation}
where we define%
\begin{equation}
\tilde{T}=Tl\text{, }\tilde{r}_{+}=r_{+}/l\text{, }\tilde{Q}=Q/l\text{,
}\tilde{a}_{i}=a_{i}l^{-c_{i}}\text{ and }\tilde{A}_{t}^{\prime}\left(
r_{+}\right)  =lA_{t}^{\prime}\left(  r_{+}\right)  \text{,}%
\end{equation}
and $\tilde{A}_{t}^{\prime}\left(  r_{+}\right)  $ is determined by%
\begin{equation}
\mathcal{L}^{\prime}\left(  \frac{\tilde{A}_{t}^{\prime}\left(  r_{+}\right)
}{2},\tilde{a}_{i}\right)  \tilde{A}_{t}^{\prime}\left(  r_{+}\right)
=\frac{\tilde{Q}}{\tilde{r}_{+}^{2}}.
\end{equation}
Solving eqn. $\left(  \ref{eq:Ttildal}\right)  $, we find that $\tilde{r}_{+}$
can be expressed as a function of $\tilde{T}$, $\tilde{Q}$ and $\tilde{a}_{i}%
$: $\tilde{r}_{+}=\tilde{r}_{+}(\tilde{T},\tilde{Q},\tilde{a}_{i})$. With
$\tilde{r}_{+}=\tilde{r}_{+}(\tilde{T},\tilde{Q},\tilde{a}_{i})$, we can
express the thermodynamic quantities in terms of $\tilde{T},\tilde{Q}$ and
$\tilde{a}_{i}$, e.g., the Gibbs free energy is given by%
\[
\tilde{F}\equiv F/l=\tilde{F}(\tilde{T},\tilde{Q},\tilde{a}_{i}).
\]

\begin{figure}[tb]
\begin{center}
\subfigure[{~\scriptsize Branches around a local minimum of $\tilde{T}=\tilde{T}_{\min}$.}]{
\includegraphics[width=0.95\textwidth]{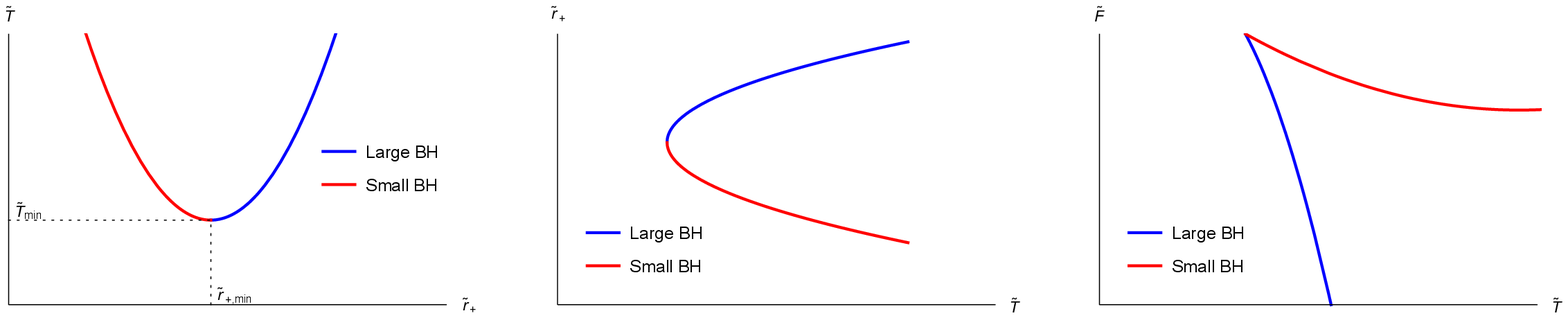}\label{fig:M:a}}
\subfigure[{~\scriptsize Branches around a local maximum of $\tilde{T}=\tilde{T}_{\max}$.}]{
\includegraphics[width=0.95\textwidth]{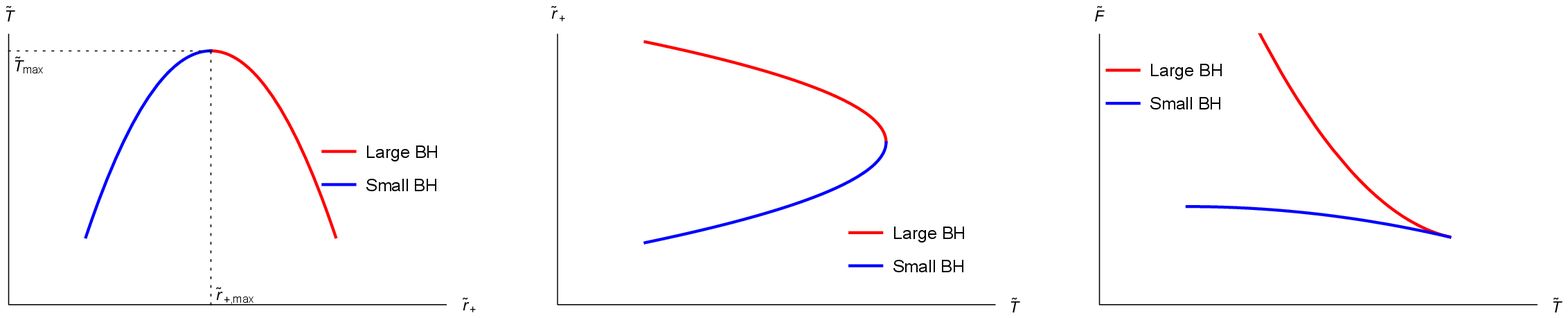}\label{fig:M:b}}
\end{center}
\caption{{\footnotesize Branches of black holes around local extremums of
$\tilde{T}=\tilde{T}_{\min}$ and $\tilde{T}=\tilde{T}_{\max}$. Right panels:
Gibbs free energy vs temperature. The blue branches are thermodynamically
preferred and thermally stable. The red ones are thermally unstable. As
$\tilde{T}\rightarrow\tilde{T}_{\max}/\tilde{T}_{\min}$, the red/blue branches
become electrically stable/unstable.}}%
\label{fig:M}%
\end{figure}

The rich phase structure of the black hole comes from solving eqn. $\left(
\ref{eq:Ttildal}\right)  $, i.e., $\tilde{T}=\tilde{T}(\tilde{r}_{+},\tilde
{Q},\tilde{a}_{i})$, for $\tilde{r}_{+}$. If $\tilde{T}(\tilde{r}_{+}%
,\tilde{Q},\tilde{a}_{i})$ is a monotonic function with respect to $\tilde
{r}_{+}$ for some values of $\tilde{Q}$ and $\tilde{a}_{i}$, there would be
only one branch for the black hole. More often, with fix $\tilde{Q}$ and
$\tilde{a}_{i}$, there exists a local minimum/maximum for $\tilde{T}(\tilde
{r}_{+},\tilde{Q},\tilde{a}_{i})$ at $\tilde{r}_{+}=\tilde{r}_{+,\min}%
/\tilde{r}_{+}=\tilde{r}_{+,\max}$. In this case, there are more than one
branch for the black hole. In FIG. \ref{fig:M:a}, we plot two branches, namely
small BH and large BH, around a local minimum of $\tilde{T}=\tilde{T}_{\min}$.
The Gibbs energy of these two branches is displayed in the right panel of FIG.
\ref{fig:M:a}. Since $\partial\tilde{F}(\tilde{T},\tilde{Q},\tilde{a}%
_{i})/\partial\tilde{T}=-16\pi^{2}\tilde{r}_{+}^{2}$, the upper branch is
small BH while lower one is large BH, which means that the large BH branch is
thermodynamically preferred. Similarly, two branches around a local maximum of
$\tilde{T}=\tilde{T}_{\max}$, small BH and large BH, are shown in FIG.
\ref{fig:M:b}. The upper/lower branch in the right panel of FIG. \ref{fig:M:b}
is large/small BH since it has more/less negative slope. So the small BH
branch is thermodynamically preferred in this case. In general, one might need
to figure out how the existence of local extremums depend on values of
$\tilde{Q}$ and $\tilde{a}_{i}$ to study the phase structure of the black hole.

After the black hole's branches are obtained, it is interesting to consider
their thermodynamic stabilities against thermal and electrical fluctuations.
In canonical ensemble, the first quantity we consider is the specific heat at
constant electric charge and pressure:%
\begin{equation}
C_{Q,P}=T\left(  \frac{\partial S}{\partial T}\right)  _{Q,P}=32l^{2}\pi
^{2}\tilde{r}_{+}\tilde{T}\frac{\partial\tilde{r}_{+}(\tilde{T},\tilde
{Q},\tilde{a}_{i})}{\partial\tilde{T}}. \label{eq:CQ}%
\end{equation}
Since the entropy is proportional to the size of the black hole, a positive
specific heat means that the black hole radiate less when it is smaller. Thus,
the thermal stability of the branch follows from $C_{Q,P}\geq0$. From eqn.
$\left(  \ref{eq:CQ}\right)  $, it shows that the large/small BH branch in
FIG. \ref{fig:M:a} and the small/large BH branch in FIG. \ref{fig:M:b} are
both thermally stable/unstable. Note that $\partial^{2}\tilde{F}(\tilde
{T},\tilde{Q},\tilde{a}_{i})/\partial^{2}\tilde{T}=-l^{-2}C_{Q,P}$, and hence
the thermally stable/unstable branches are concave downward/upward in right
panels of FIG. \ref{fig:M}.

The second quantity is
\begin{equation}
\epsilon_{T}=\left(  \frac{\partial Q}{\partial\Phi}\right)  _{T}=l\left[
\frac{\partial\Phi(\tilde{r}_{+},\tilde{Q},\tilde{a}_{i})}{\partial\tilde{Q}%
}-\frac{\partial\Phi(\tilde{r}_{+},\tilde{Q},\tilde{a}_{i})}{\partial\tilde
{r}_{+}}\frac{\frac{\partial\tilde{T}(\tilde{r}_{+},\tilde{Q},\tilde{a}_{i}%
)}{\partial Q}}{\frac{\partial\tilde{T}(\tilde{r}_{+},\tilde{Q},\tilde{a}%
_{i})}{\partial\tilde{r}_{+}}}\right]  ^{-1}, \label{eq:epsilon}%
\end{equation}
which describes how the black hole's electrostatic potential respond to its
charge. Here, $\tilde{r}_{+}$ is understood as $\tilde{r}_{+}(\tilde{T}%
,\tilde{Q},\tilde{a}_{i})$. For a positive value of $\epsilon_{T}$, as more
charges are placed on the black hole, its potential increases and hence make
it harder to move the system from equilibrium. The electrical stability of the
branch then follows from $\epsilon_{T}\geq0$. For the potential $\Phi$, it is
natural to expect that $\partial\Phi(\tilde{r}_{+},\tilde{Q},\tilde{a}%
_{i})/\partial\tilde{r}_{+}<0$. We also assume $Q>0$ and hence $\tilde{A}%
_{t}^{\prime}\left(  r_{+}\right)  >0$, which gives%
\begin{equation}
\frac{\partial\tilde{T}(\tilde{r}_{+},\tilde{Q},\tilde{a}_{i})}{\partial
Q}=-\frac{1}{8\pi\tilde{r}_{+}}\tilde{A}_{t}^{\prime}\left(  r_{+}\right)  <0.
\end{equation}
Eqn. $\left(  \ref{eq:epsilon}\right)  $ shows that in FIG. \ref{fig:M:a},
$\varepsilon_{T}^{-1}\rightarrow+\infty$ for the small BH branch and
$\varepsilon_{T}^{-1}\rightarrow-\infty$ for the large BH branch as $\tilde
{T}\rightarrow\tilde{T}_{\min}$. Similarly in FIG. \ref{fig:M:b},
$\varepsilon_{T}^{-1}\rightarrow-\infty$ for the small BH branch and
$\varepsilon_{T}^{-1}\rightarrow+\infty$ for the large BH branch as $\tilde
{T}\rightarrow\tilde{T}_{\max}$.

Finally, we turn to the critical point, which is an inflection point and
obtained by%
\begin{equation}
\frac{\partial\tilde{T}(\tilde{r}_{+},\tilde{Q},\tilde{a}_{i})}{\partial
\tilde{r}_{+}}=0\text{ and }\frac{\partial^{2}\tilde{T}(\tilde{r}_{+}%
,\tilde{Q},\tilde{a}_{i})}{\partial\tilde{r}_{+}^{2}}=0.
\end{equation}
Solving the above equations gives%
\begin{equation}
\tilde{r}_{+,c}=\tilde{r}_{+,c}\left(  \tilde{a}_{i}\right)  \text{, }%
\tilde{Q}_{c}=\tilde{Q}_{c}\left(  \tilde{a}_{i}\right)  \text{ and }\tilde
{T}_{c}=\tilde{T}_{c}\left(  \tilde{a}_{i}\right)  .
\end{equation}
Defining the specific volume $v=\frac{r_{+}}{8\pi}$
\cite{IN-Gunasekaran:2012dq}, one finds that%

\begin{equation}
\rho_{c}=\frac{P_{c}v_{c}}{T_{c}}=\frac{3}{4\pi}\frac{\tilde{r}_{+,c}\left(
\tilde{a}_{i}\right)  }{\tilde{T}_{c}\left(  \tilde{a}_{i}\right)  }.
\end{equation}
Consider the case in which $\mathcal{L}\left(  s,a_{i}\right)  $ is a power
series expansion of $s$:%
\begin{equation}
\mathcal{L}\left(  s,a_{i}\right)  =s+\frac{a_{1}}{2}s^{2}+\frac{a_{2}}%
{3}s^{3}+\cdots.
\end{equation}
For small values of $a_{i}$, we find%
\begin{align}
\tilde{Q}_{c}  &  =\frac{1}{3}+\frac{7}{18}\tilde{a}_{1}+\frac{11\left(
\tilde{a}_{1}^{2}+16\tilde{a}_{2}\right)  }{216}+\cdots,\nonumber\\
\tilde{T}_{c}  &  =\frac{1}{\pi}\sqrt{\frac{2}{3}}-\frac{\tilde{a}_{1}}%
{3\sqrt{6}\pi}-\frac{9\tilde{a}_{1}^{2}+32\tilde{a}_{2}}{36\sqrt{6}\pi}%
+\cdots,\nonumber\\
\tilde{r}_{+,c}  &  =\frac{1}{\sqrt{6}}-\frac{7\tilde{a}_{1}}{6\sqrt{6}}%
+\frac{27\tilde{a}_{1}^{2}-352\tilde{a}_{2}}{72\sqrt{6}}+\cdots,\\
\rho_{c}  &  =\frac{P_{c}v_{c}}{T_{c}}=\frac{3\left(  1-\tilde{a}_{1}\right)
}{8}+\frac{3\tilde{a}_{1}^{2}-40\tilde{a}_{2}}{24}+\cdots.\nonumber
\end{align}
where%
\[
\tilde{Q}_{c}=Q_{c}\sqrt{P_{c}/6}\text{, }\tilde{T}_{c}=T_{c}\sqrt{6/P_{c}%
}\text{, }\tilde{r}_{+,c}=r_{+,c}\sqrt{P_{c}/6}\text{ and }\tilde{a}%
_{i}=\left(  P_{c}/6\right)  ^{i}a_{i}.
\]
The leading value of $\rho_{c}$ is $3/8$, which reproduces the critical value
$\rho_{c}$ of RN-AdS black holes.

\section{Born-Infeld AdS Black Hole}

\label{Sec:BIABH}

Born-Infeld electrodynamics is described by the Lagrangian density%
\begin{equation}
\mathcal{L}\left(  s\right)  =\frac{1}{a}\left(  1-\sqrt{1-2as}\right)
\text{,} \label{eq:BI}%
\end{equation}
where the coupling parameter $a$ is related to the string tension
$\alpha^{\prime}$ as $a=\left(  2\pi\alpha^{\prime}\right)  ^{2}>0$. When
$a=0$, we can recover the Maxwell Lagrangian. Solving eqn. $\left(
\ref{eq:QAt}\right)  $ for $A_{t}^{\prime}\left(  r\right)  $ gives
\begin{equation}
A_{t}^{\prime}\left(  r\right)  =\frac{Q}{\sqrt{r^{4}+aQ^{2}}}.
\end{equation}
It follows that the potential of the black hole is%
\begin{equation}
\Phi=\frac{4\pi Q}{r_{+}}\text{ }_{2}F_{1}\left(  \frac{1}{4},\frac{1}%
{2},\frac{5}{4};-\frac{aQ^{2}}{r_{+}^{4}}\right)  ,
\end{equation}
where $_{2}F_{1}\left(  a,b,c;x\right)  $ is the hypergeometric function.

The equation of state $\left(  \ref{eq:Ttildal}\right)  $ becomes
\begin{equation}
\tilde{T}\left(  \tilde{r}_{+}\right)  \equiv\frac{h\left(  \tilde{r}%
_{+}\right)  }{4\pi\tilde{r}_{+}}=\frac{1}{4\pi\tilde{r}_{+}}\left(
1+3\tilde{r}_{+}^{2}-\frac{1}{2}\frac{\tilde{Q}^{2}}{\tilde{r}_{+}^{2}%
+\sqrt{\tilde{r}_{+}^{4}+\tilde{a}\tilde{Q}^{2}}}\right)  , \label{eq:BIT}%
\end{equation}
where $\tilde{a}=a/l^{2}$, and we define $h\left(  \tilde{r}_{+}\right)  $ for
later use. Noting that $h\left(  0\right)  =1-\frac{\tilde{Q}}{2\sqrt
{\tilde{a}}}$, $h\left(  \infty\right)  \rightarrow+\infty$ and $h\left(
\tilde{r}_{+}\right)  $ is a strictly increasing function, one finds that
\begin{align*}
Q^{2}  &  \geq4a\text{: }\tilde{T}\left(  \tilde{r}_{+}\right)  =0\text{ has
only one solution }\tilde{r}_{+}=\tilde{r}_{e}\geq0,\\
Q^{2}  &  <4a\text{: }\tilde{T}\left(  \tilde{r}_{+}\right)  >0\text{ for
}\tilde{r}_{+}\geq0\text{,}%
\end{align*}
where $\tilde{r}_{e}$ corresponds to an extremal black hole.

To study behavior of local extremums of $\tilde{T}\left(  \tilde{r}%
_{+}\right)  $, we consider the equation $\tilde{T}^{\prime\prime}\left(
\tilde{r}_{+}\right)  =0$, which becomes
\begin{equation}
z\left(  x\right)  \equiv x^{3}-\frac{3\tilde{Q}^{2}}{2}x^{2}+\tilde{a}%
\tilde{Q}^{4}=0, \label{eq:z(x)}%
\end{equation}
with $x=\sqrt{\tilde{a}\tilde{Q}^{2}+\tilde{r}_{+}^{4}}$. Since $\lim
_{x\rightarrow\pm\infty}z\left(  x\right)  =\pm\infty$, $z^{\prime}\left(
0\right)  =0$ and $z^{\prime}(\tilde{Q}^{2})=0$, $z\left(  x\right)  $ has a
local maximum of $z\left(  0\right)  =\tilde{a}\tilde{Q}^{4}>0$ at $x=0$ and a
local minimum of $z(\tilde{Q}^{2})=(-\tilde{Q}^{2}/2+\tilde{a})\tilde{Q}^{4}$
at $x=\tilde{Q}^{2}$. If the local minimum is not greater than zero $\left(
Q^{2}\geq2a\right)  $, there are two positive real roots $x_{1}\geq\tilde
{Q}^{2}\geq x_{2}>0$ to the equation $\left(  \ref{eq:z(x)}\right)  $.
Otherwise $\left(  Q^{2}<2a\right)  $, this equation has no positive real
roots. To make $\tilde{r}_{+}=(x^{2}-\tilde{a}\tilde{Q}^{2})^{1/4}$ real, we
also require $x\geq\sqrt{\tilde{a}}\tilde{Q}$. For $x_{1}$, one always has
that $x_{1}\geq\tilde{Q}^{2}>\sqrt{\tilde{a}}\tilde{Q}$ since $Q^{2}\geq2a$.
To have $x_{2}\geq\sqrt{\tilde{a}}\tilde{Q}$, we need to have $z(\sqrt
{\tilde{a}}\tilde{Q})\leq0\Rightarrow Q^{2}$ $\geq4a$. With solutions of
$\tilde{T}^{\prime\prime}\left(  \tilde{r}_{+}\right)  =0$, it is easy to
analyze the existence of the local extremums of $\tilde{T}^{\prime}\left(
\tilde{r}_{+}\right)  $, results of which are summarized in Table \ref{tab:1}.

\begin{table}[tbh]
\centering%
\begin{tabular}
[c]{|c|c|c|c|c|}\hline
& {\footnotesize $\tilde{T}^{\prime}\left(  0\right)  $} &
{\footnotesize $\tilde{T}^{\prime}\left(  +\infty\right)  $} &
{\footnotesize Solution of $\tilde{T}^{\prime\prime}\left(  \tilde{r}%
_{+}\right)  =0$} & {\footnotesize Extremums of $\tilde{T}^{\prime}\left(
\tilde{r}_{+}\right)  $}\\\hline
{\footnotesize $Q^{2}>4a$} & {\footnotesize $+\infty$} &
{\footnotesize $3/4\pi$} & {\footnotesize $\tilde{r}_{1}>0$} &
{\footnotesize Minimum at $\tilde{r}_{+}=\tilde{r}_{1}$}\\\hline
{\footnotesize $Q^{2}=4a$} & {\footnotesize $>0$} & {\footnotesize $3/4\pi$} &
{\footnotesize $\tilde{r}_{1}>0$, $0$} & {\footnotesize Minimum at $\tilde
{r}_{+}=\tilde{r}_{1}$}\\\hline
{\footnotesize $4a>Q^{2}>2a$} & {\footnotesize $-\infty$} &
{\footnotesize $3/4\pi$} & {\footnotesize $\tilde{r}_{1}>\tilde{r}_{2}>0$} &
{\footnotesize Minimum/Maximum at $\tilde{r}_{+}=\tilde{r}_{1}/\tilde{r}_{2}$%
}\\\hline
{\footnotesize $Q^{2}=2a$} & {\footnotesize $-\infty$} &
{\footnotesize $3/4\pi$} & {\footnotesize $\tilde{r}_{1}>0$, $\tilde
{T}^{\prime\prime}\left(  \tilde{r}_{+}\right)  \geq0$} & {\footnotesize None}%
\\\hline
{\footnotesize $Q^{2}<2a$} & {\footnotesize $-\infty$} &
{\footnotesize $3/4\pi$} & {\footnotesize None, $\tilde{T}^{\prime\prime
}\left(  \tilde{r}_{+}\right)  >0$} & {\footnotesize None}\\\hline
\end{tabular}
\caption{{\small Solution of $\tilde{T}^{\prime\prime}\left(  \tilde{r}%
_{+}\right)  =0$ and the local extremums of $\tilde{T}^{\prime}\left(
\tilde{r}_{+}\right)  $ in various cases, where $\tilde{r}_{i}=\left(
x_{i}^{2}-\tilde{a}\tilde{Q}^{2}\right)  ^{1/4}$.}}%
\label{tab:1}%
\end{table}

When solving eqn. $\left(  \ref{eq:BIT}\right)  $ for $\tilde{r}_{+}$ in terms
of $\tilde{T}$, the solution $\tilde{r}_{+}(\tilde{T})$ is often a multivalued
function. The parameters $\tilde{a}$ and $\tilde{Q}$ determine the number of
the branches of $\tilde{r}_{+}(\tilde{T})$ and the phase structure of the
black hole. In what follows, we find six regions in the $\tilde{a}$-$\tilde
{Q}$ plane, in each of which the black hole has the distinct behavior of the
branches and the phase structure:

\begin{figure}[tbh]
\begin{center}
\subfigure[{~\scriptsize Region I: $a/l^{2}=0.01$ and $Q/l=0.4$. There is no phase transition.}]{
\includegraphics[width=1.01\textwidth]{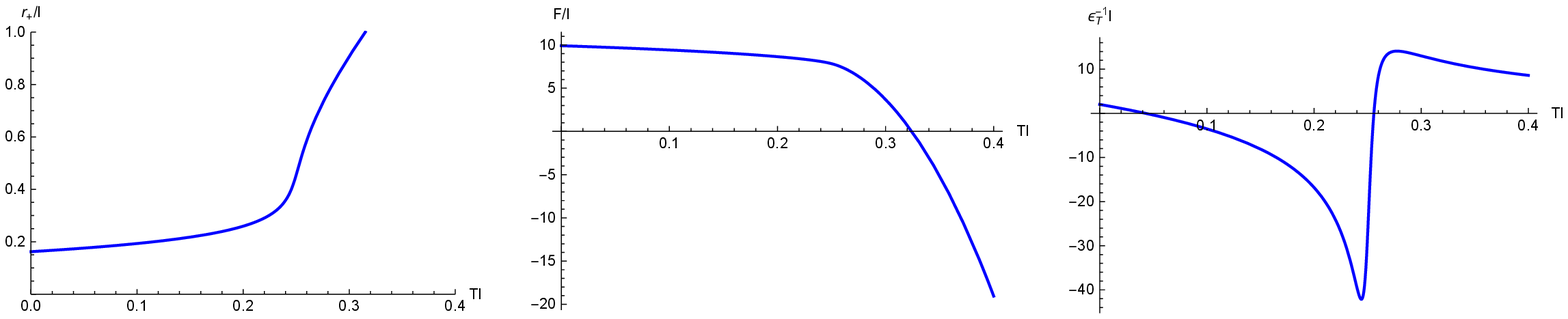}\label{fig:DBIR12:a}}
\subfigure[{~\scriptsize Region II: $a/l^{2}=0.01$ and $Q/l=0.25$. There is a first order phase transition between small BH and large BH.}]{
\includegraphics[width=1.01\textwidth]{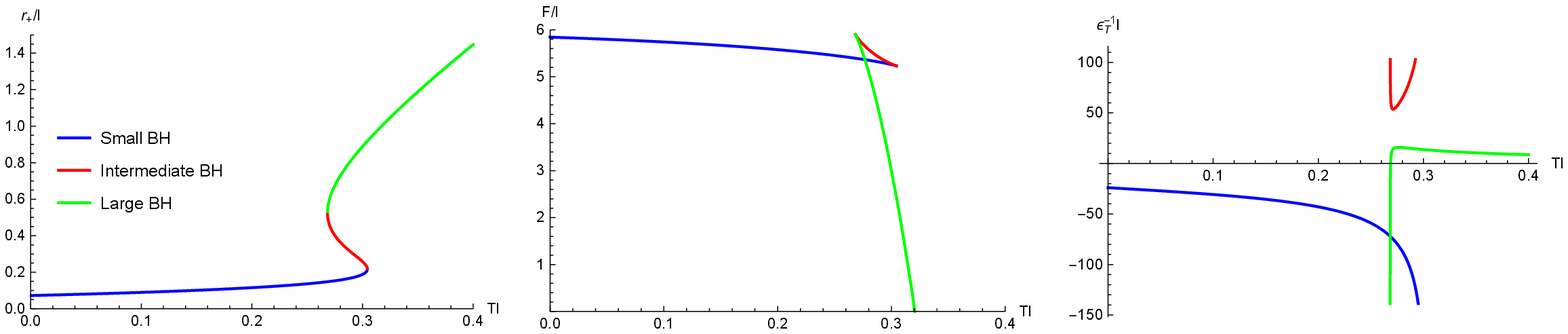}\label{fig:DBIR12:b}}
\end{center}
\caption{{\footnotesize Plot of $\tilde{r}_{+}$, $\tilde{F}$ and $\epsilon
_{T}^{-1}l$ against $\tilde{T}$ for BI-AdS black holes in Regions I and II.
Black holes in these regions are RN type since there exist extremal black hole
solutions. Regions I and II can be considered as reminiscent of RN-AdS black
holes. The blue and green branches are thermally stable.}}%
\label{fig:DBIR12}%
\end{figure}

\begin{figure}[ptb]
\begin{center}
\subfigure[{~\scriptsize Region III: $a/l^{2}=0.01$ and $Q/l=0.195$. There is a first order phase transition between small BH and large BH.}]{
\includegraphics[width=1\textwidth]{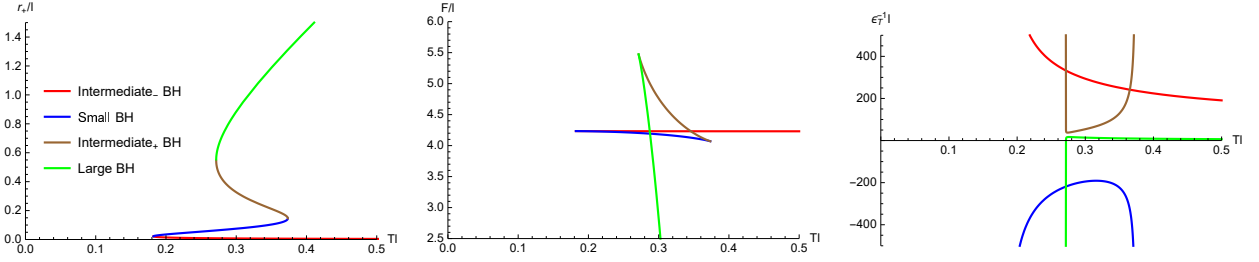}\label{fig:DBIR36:a}}
\subfigure[{~ \scriptsize Region IV: $a/l^{2}=0.01$ and $Q/l=0.188$.  The arrows in the inset
indicate increasing $\tilde{T}$. As $\tilde{T}$ increases, the black hole
jumps from the large BH branch to the small BH one, corresponding to the
zeroth order phase transition between small BH and large BH. Further
increasing $\tilde{T}$, there would be a first order phase transition
returning to large BH. Here we observe
LBH/SBH/LBH reentrant phase transition.}]{
\includegraphics[width=1\textwidth]{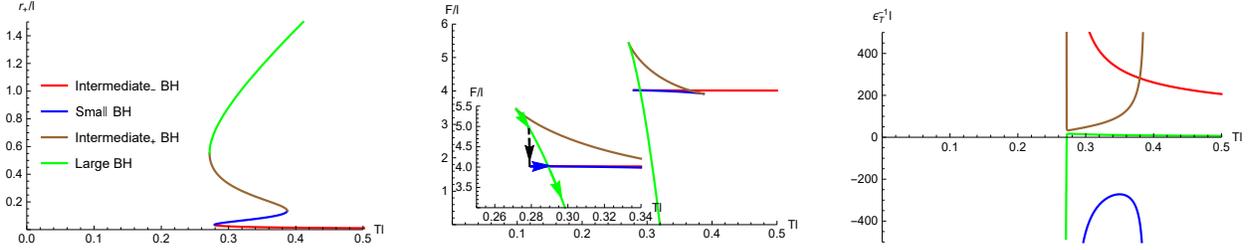}\label{fig:DBIR36:b}}
\subfigure[{~ \scriptsize Region V: $a/l^{2}=0.01$ and $Q/l=0.185$. There is no phase transition.}]{
\includegraphics[width=1\textwidth]{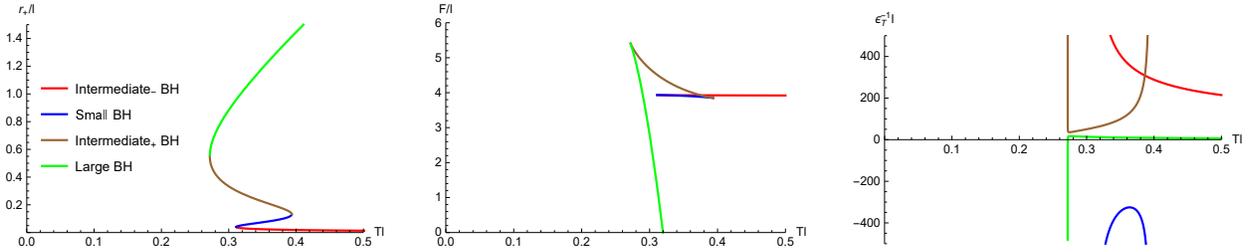}\label{fig:DBIR36:c}}
\subfigure[{~ \scriptsize Region VI: $a/l^{2}=0.01$ and $Q/l=0.15$. There is no phase transition.}]{
\includegraphics[width=1\textwidth]{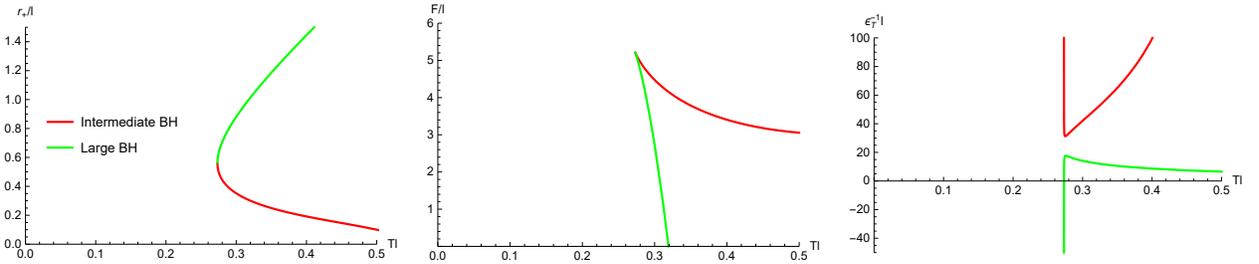}\label{fig:DBIR36:d}}
\end{center}
\caption{{\footnotesize Plot of $\tilde{r}_{+}$, $\tilde{F}$ and $\epsilon
_{T}^{-1}l$ against $\tilde{T}$ for BI-AdS black holes in Regions III, IV,V
and VI. Black holes in these regions are Schwarzschild-like\ type\ since they
only exist for large enough $\tilde{T}$. The blue and green branches are
always thermally stable. Small BH is electrically unstable while large BH is
almost electrically stable.}}%
\end{figure}

\begin{itemize}
\item Region I: $Q^{2}\geq4a$ and $\tilde{T}^{\prime}\left(  \tilde{r}%
_{1}\right)  \geq0$. In this region, $\tilde{T}^{\prime}\left(  \tilde{r}%
_{+}\right)  \geq\tilde{T}^{\prime}\left(  \tilde{r}_{1}\right)  \geq0$ and
hence $\tilde{T}\left(  \tilde{r}_{+}\right)  $ is\ an increasing function. So
there is only one branch for $\tilde{r}_{+}(\tilde{T})$, which is thermally
stable. Since $\tilde{T}\left(  \tilde{r}_{+}\right)  =0$ has a solution in
this region, this branch can extend to zero temperature. For a black hole with
$\tilde{a}=0.01$ and $\tilde{Q}=0.4$ in this region, we plot the radius
$\tilde{r}_{+}$, the Gibbs energy $\tilde{F}$ and the isothermal permittivity
$\epsilon_{T}^{-1}l$ as functions of $\tilde{T}$ in FIG. \ref{fig:DBIR12:a},
which shows that this black hole is electrically stable for small enough and
large enough $\tilde{T}$. However for large enough $\tilde{a}$, the black hole
is always electrically stable.

\item Region II: $Q^{2}\geq4a$ and $\tilde{T}^{\prime}\left(  \tilde{r}%
_{1}\right)  <0$. In this region, $\tilde{T}^{\prime}\left(  \tilde{r}%
_{+}\right)  =0$ has two solutions $\tilde{r}_{+}=$\ $\tilde{r}_{\text{max}}$
and\ $\tilde{r}_{\text{min}}$ with $\tilde{r}_{\text{max}}<\tilde{r}%
_{1}<\tilde{r}_{\text{min}}$. Since $\tilde{T}\left(  +\infty\right)
=+\infty$, $\tilde{T}\left(  \tilde{r}_{+}\right)  $ has a local maximum of
$\tilde{T}_{\text{max}}=\tilde{T}\left(  \tilde{r}_{\text{max}}\right)  $ at
$\tilde{r}_{+}=$\ $\tilde{r}_{\text{max}}$ and a local minimum of $\tilde
{T}_{\text{min}}=\tilde{T}\left(  \tilde{r}_{\text{min}}\right)  $ at
$\tilde{r}_{+}=$\ $\tilde{r}_{\text{min}}$. There are three branches for
$\tilde{r}_{+}(\tilde{T})$: small BH for $0\leq\tilde{T}\leq\tilde
{T}_{\text{max}}$, intermediate BH for $\tilde{T}_{\text{min}}\leq\tilde
{T}\leq\tilde{T}_{\text{max}}$ and large BH for $\tilde{T}\geq$ $\tilde
{T}_{\text{min}}$, which are displayed in the left panel of FIG.
\ref{fig:DBIR12:b}. The Gibbs free energy of the three branches is plotted in
the middle panel, which shows that there is a first order phase transition
between small BH and large BH occurring at $\tilde{T}=\tilde{T}_{\text{first}%
}$ with $\tilde{T}_{\text{min}}\leq\tilde{T}_{\text{first}}\leq\tilde
{T}_{\text{max}}$. Both the small BH and large BH branches are thermally
stable. As explained in section \ref{Sec:NBH}, $\epsilon_{T}^{-1}$ of small BH
goes to $-\infty$ as $\tilde{T}\rightarrow\tilde{T}_{\text{max}}$ while that
of large BH goes to $+\infty$ as $\tilde{T}\rightarrow\tilde{T}_{\text{min}}$.
The right panel shows that small BH is electrically unstable while large BH is
almost electrically stable.

\item Region III: $4a>Q^{2}>2a$, $\tilde{T}^{\prime}\left(  \tilde{r}%
_{1}\right)  <0$, $\tilde{T}^{\prime}\left(  \tilde{r}_{2}\right)  >0$ and
$\tilde{T}\left(  \tilde{r}_{\text{min2}}\right)  <\tilde{T}\left(  \tilde
{r}_{\text{min1}}\right)  $. In this region, $\tilde{T}^{\prime}\left(
\tilde{r}_{+}\right)  =0$ has three solutions $\tilde{r}_{+}=$\ $\tilde
{r}_{\text{max}}$, $\tilde{r}_{\text{min1}}$ and $\tilde{r}_{\text{min2}}$
with\ $\tilde{r}_{\text{min2}}<\tilde{r}_{2}<\tilde{r}_{\text{max}}<\tilde
{r}_{1}<\tilde{r}_{\text{min1}}$. So $\tilde{T}\left(  \tilde{r}_{+}\right)  $
has a local maximum of $\tilde{T}_{\text{max}}=\tilde{T}\left(  \tilde
{r}_{\text{max}}\right)  $ at $\tilde{r}_{+}=$\ $\tilde{r}_{\text{max}}$, a
local minimum of $\tilde{T}_{\text{min1}}=\tilde{T}\left(  \tilde
{r}_{\text{min1}}\right)  $ at $\tilde{r}_{+}=$\ $\tilde{r}_{\text{min1}}$ and
a global minimum of $\tilde{T}_{\text{min2}}=\tilde{T}\left(  \tilde
{r}_{\text{min2}}\right)  $ at $\tilde{r}_{+}=$\ $\tilde{r}_{\text{min2}}$.
There are four branches for $\tilde{r}_{+}(\tilde{T})$: intermediate$_{-}$ BH
for $\tilde{T}\geq\tilde{T}_{\text{min2}}$, small BH for $\tilde
{T}_{\text{min2}}\leq\tilde{T}\leq\tilde{T}_{\text{max}}$, intermediate$_{+}$
BH for $\tilde{T}_{\text{min1}}\leq\tilde{T}\leq\tilde{T}_{\text{max}}$ and
large BH for $\tilde{T}\geq\tilde{T}_{\text{min1}}$, which are displayed in
the left panel of FIG. \ref{fig:DBIR36:a}. Note that there is no black hole
solution when $\tilde{T}<\tilde{T}_{\text{min2}}$. The Gibbs free energy of
the four branches is plotted in the middle panel, which shows that there is a
first order phase transition between small BH and large BH occurring at
$\tilde{T}=\tilde{T}_{\text{first}}$ with $\tilde{T}_{\text{min2}}\leq
\tilde{T}_{\text{first}}\leq\tilde{T}_{\text{max}}$. Both the small BH and
large BH branches are thermally stable, while intermediate$_{\pm}$ BH branches
are not. Similarly to Region II, the right panel shows that small BH is
electrically unstable while large BH is almost electrically stable.

\item Region IV: $4a>Q^{2}>2a$, $\tilde{T}^{\prime}\left(  \tilde{r}%
_{1}\right)  <0$, $\tilde{T}^{\prime}\left(  \tilde{r}_{2}\right)  >0$,
$\tilde{T}_{\text{min2}}\geq\tilde{T}_{\text{min1}}$ and $\tilde{F}_{S}%
(\tilde{T}_{\text{min2}})<\tilde{F}_{L}(\tilde{T}_{\text{min2}})$, where
$\tilde{F}_{S/L}$ is the Gibbs free energy of the small/large BH branch. In
this region, $\tilde{T}\left(  \tilde{r}_{+}\right)  $ has a local maximum of
$\tilde{T}_{\text{max}}=\tilde{T}\left(  \tilde{r}_{\text{max}}\right)  $ at
$\tilde{r}_{+}=$\ $\tilde{r}_{\text{max}}$, a local minimum of $\tilde
{T}_{\text{min2}}=\tilde{T}\left(  \tilde{r}_{\text{min2}}\right)  $ at
$\tilde{r}_{+}=$\ $\tilde{r}_{\text{min2}}$ and a global minimum of $\tilde
{T}_{\text{min1}}=\tilde{T}\left(  \tilde{r}_{\text{min1}}\right)  $ at
$\tilde{r}_{+}=$\ $\tilde{r}_{\text{min1}}$. There are four branches for
$\tilde{r}_{+}(\tilde{T})$: intermediate$_{-}$ BH for $\tilde{T}\geq\tilde
{T}_{\text{min2}}$, small BH for $\tilde{T}_{\text{min2}}\leq\tilde{T}%
\leq\tilde{T}_{\text{max}}$, intermediate$_{+}$ BH for $\tilde{T}%
_{\text{min1}}\leq\tilde{T}\leq\tilde{T}_{\text{max}}$ and large BH for
$\tilde{T}\geq\tilde{T}_{\text{min1}}$, which are displayed in the left panel
of FIG. \ref{fig:DBIR36:b}. The Gibbs free energy of the four branches is
plotted in the middle panel. As $\tilde{T}$ increases from $\tilde
{T}_{\text{min1}}$, the black hole follows direction of arrows in the inset.
It shows that there is a finite jump in Gibbs free energy leading to a zeroth
order phase transition from large BH to small BH, followed by a first order
phase transition returning to large BH. This LBH/SBH/LBH transition
corresponds to a reentrant phase transition.

\item Region V: $4a>Q^{2}>2a$, $\tilde{T}^{\prime}\left(  \tilde{r}%
_{1}\right)  <0$, $\tilde{T}^{\prime}\left(  \tilde{r}_{2}\right)  >0$,
$\tilde{T}_{\text{min2}}>\tilde{T}_{\text{min1}}$ and $\tilde{F}_{S}(\tilde
{T}_{\text{min2}})\geq\tilde{F}_{L}(\tilde{T}_{\text{min2}})$. As shown in the
left panel of FIG. \ref{fig:DBIR36:c}, the four branches of $\tilde{r}%
_{+}(\tilde{T})$ in this region are the same as in Region IV. However, the
middle panel shows that the large BH branch is always thermodynamically
preferred for $\tilde{T}\geq\tilde{T}_{\text{min1}}$, and hence there is no
phase transition in this region.

\item Region VI: $4a>Q^{2}>2a$ and $\tilde{T}^{\prime}\left(  \tilde{r}%
_{1}\right)  >0$ or $\tilde{T}^{\prime}\left(  \tilde{r}_{2}\right)  <0$; or
$Q^{2}<2a$. It can show that $\tilde{T}^{\prime}\left(  \tilde{r}_{+}\right)
=0$ has only one solution $\tilde{r}_{+}=\tilde{r}_{\text{min}}$ in this
region. Since $\tilde{T}\left(  +\infty\right)  =+\infty$, $\tilde{T}\left(
\tilde{r}_{+}\right)  $ has a global minimum of $\tilde{T}_{\text{min}}%
=\tilde{T}\left(  \tilde{r}_{\text{min}}\right)  $ at $\tilde{r}_{+}%
=$\ $\tilde{r}_{\text{min}}$. As shown in the left panel of FIG.
\ref{fig:DBIR36:d}, there are two branches for $\tilde{r}_{+}(\tilde{T})$:
large BH and intermediate BH for $\tilde{T}\geq\tilde{T}_{\text{min}}$. The
middle panel shows that the large BH branch is always thermodynamically
preferred for $\tilde{T}\geq\tilde{T}_{\text{min}}$, and hence there is no
phase transition in this region. This region is similar to the
Schwarzschild-AdS case. The large BH branch is thermally stable and almost
electrically stable.
\end{itemize}

\begin{figure}[tb]
\begin{center}
\subfigure[{~ \scriptsize The six regions in the $\tilde{a}$-$\tilde{Q}$ plane, each of which possesses the
distinct behavior of the branches and the phase structure for BI-AdS
black holes. The LBH/SBH/LBH reentrant phase transition occurs in Region IV.
The LBH/SBH first order phase transition occurs in Regions II and III. No
phase transitions occur in Regions I,V and VI.}]{
\includegraphics[width=0.48\textwidth]{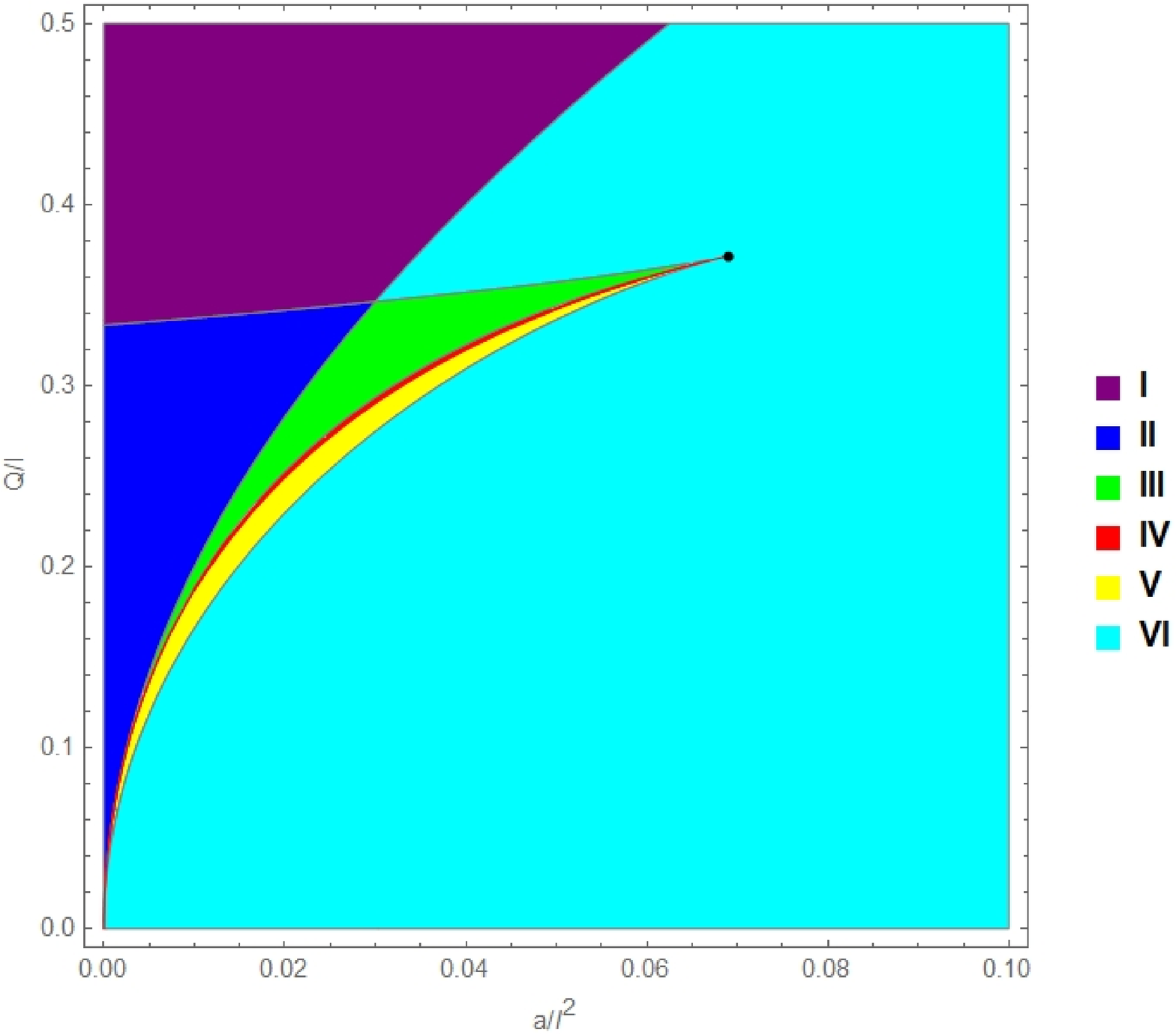}\label{fig:DBIQa:a}}
\subfigure[{~ \scriptsize The critical line has two branches, and the physical/unphysical one is
depicted by the blue/red line. $\tilde{Q}_{l}\left(  \tilde{a}\right)$ is plotted for various values of
$Q/\sqrt{a}$. There is no critical point for $Q/\sqrt{a}<\sqrt{2}$, while
there exists one physical critical point for $Q/\sqrt{a}>\sqrt{2}$. The black
hole also possesses an unphysical critical point for $\sqrt{2}<Q/\sqrt{a}<1.6948$.}]{
\includegraphics[width=0.4\textwidth]{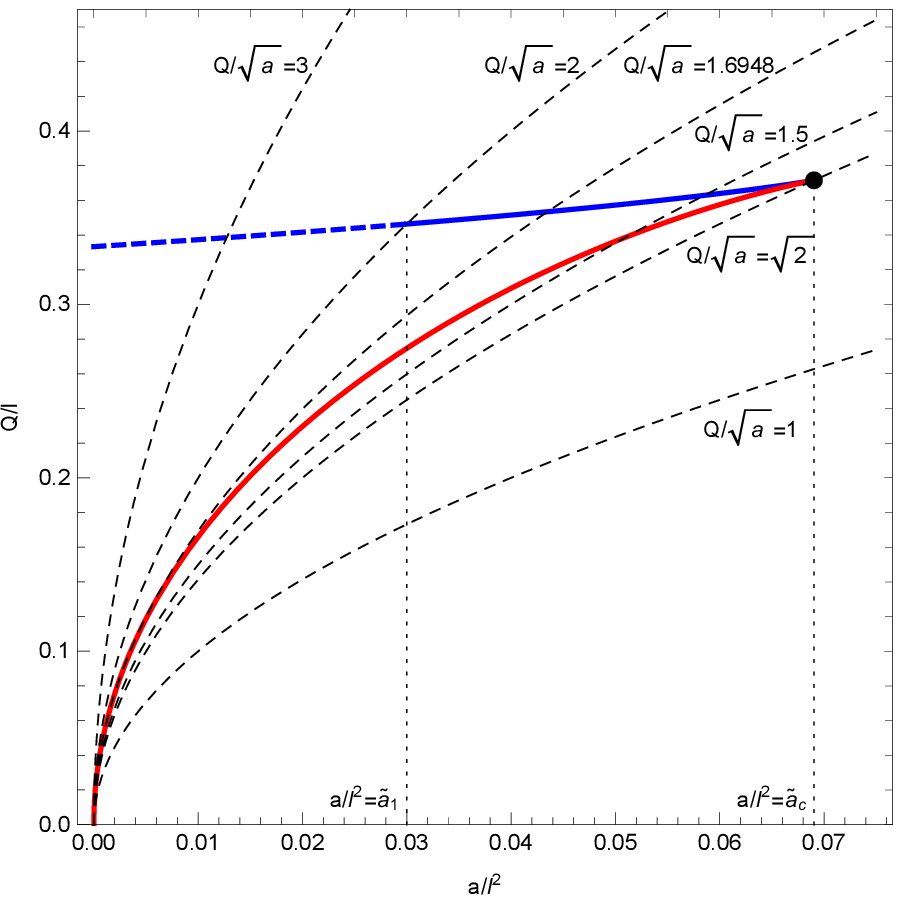}\label{fig:DBIQa:b}}
\end{center}
\caption{{\footnotesize The six regions and the critical line in the
$\tilde{a}$-$\tilde{Q}$ plane for BI-AdS black holes. The critical line
consists of $\tilde{Q}_{12}\left(  \tilde{a}\right)  $ (the blue dashed line),
$\tilde{Q}_{36}\left(  \tilde{a}\right)  $ (the blue solid line) and
$\tilde{Q}_{56}\left(  \tilde{a}\right)  $ (the red line), where $\tilde
{Q}_{ij}\left(  \tilde{a}\right)  $ is the boundary Region $i$ and Region $j$.
With fixed $Q$ and $a$, the black hole moves along the curve $\tilde{Q}%
_{l}\left(  \tilde{a}\right)  =\frac{Q}{\sqrt{a}}\sqrt{\tilde{a}}$ by varying
$P$.}}%
\label{fig:DBIQa}%
\end{figure}

In FIG. \ref{fig:DBIQa:a}, we plot these six regions in the $\tilde{a}%
$-$\tilde{Q}$ plane. It is interesting to note that the boundary between the
region in which $\tilde{T}\left(  \tilde{r}_{+}\right)  $ has $n$ extremums
and that in which $\tilde{T}\left(  \tilde{r}_{+}\right)  $ has $n+2$
extremums is the critical line, determined by%
\begin{equation}
\frac{\partial\tilde{T}(\tilde{r}_{+},\tilde{Q},\tilde{a})}{\partial\tilde
{r}_{+}}=0\text{ and }\frac{\partial^{2}\tilde{T}(\tilde{r}_{+},\tilde
{Q},\tilde{a})}{\partial\tilde{r}_{+}^{2}}=0. \label{eq:DBIcritical}%
\end{equation}
There are 3 such boundaries in FIG. \ref{fig:DBIQa:a}, i.e., $\tilde{Q}%
_{12}\left(  \tilde{a}\right)  $, $\tilde{Q}_{36}\left(  \tilde{a}\right)  $,
$\tilde{Q}_{56}\left(  \tilde{a}\right)  $, where $\tilde{Q}_{ij}\left(
\tilde{a}\right)  $ is the boundary Region $i$ and Region $j$. As shown in
FIG. \ref{fig:DBIQa:a}, the critical line has two branches: $\tilde{Q}%
_{c1}\left(  \tilde{a}\right)  =\left\{  \tilde{Q}_{12}\left(  \tilde
{a}\right)  ,\tilde{Q}_{36}\left(  \tilde{a}\right)  \right\}  $ and
$\tilde{Q}_{c2}\left(  \tilde{a}\right)  =\tilde{Q}_{56}\left(  \tilde
{a}\right)  $. We plot these two branches of the critical line in FIG.
\ref{fig:DBIQa:b}, where $\tilde{Q}_{c1}\left(  \tilde{a}\right)  $ is the
blue line, and $\tilde{Q}_{c2}\left(  \tilde{a}\right)  $ is the red line.
Note that $\tilde{Q}_{c1}\left(  \tilde{a}\right)  $ and $\tilde{Q}%
_{c2}\left(  \tilde{a}\right)  $ meet and terminate at $\left\{  \tilde{a}%
_{c},\tilde{Q}_{c}\right\}  \simeq\left\{  0.069,0.37\right\}  $, which is
represented by the black point in FIG. \ref{fig:DBIQa}. However, the middle
panel of FIG. \ref{fig:DBIR36:c} shows that the branch $\tilde{Q}_{c2}\left(
\tilde{a}\right)  $ is not physical since it does not globally minimize the
Gibbs free energy. So the critical line has only one physical branch,
$\tilde{Q}_{c1}\left(  \tilde{a}\right)  $, which is marked by the blue line.
For $\tilde{a}\leq\tilde{a}_{1}\simeq0.030$, $\tilde{Q}_{c1}\left(  \tilde
{a}\right)  $ is $\tilde{Q}_{12}\left(  \tilde{a}\right)  $ and depicted by
the blue dashed line in FIG. \ref{fig:DBIQa:b}. This part of $\tilde{Q}%
_{c1}\left(  \tilde{a}\right)  $ is reminiscent of RN-AdS black holes.

\begin{figure}[tb]
\begin{center}
\includegraphics[width=0.48\textwidth]{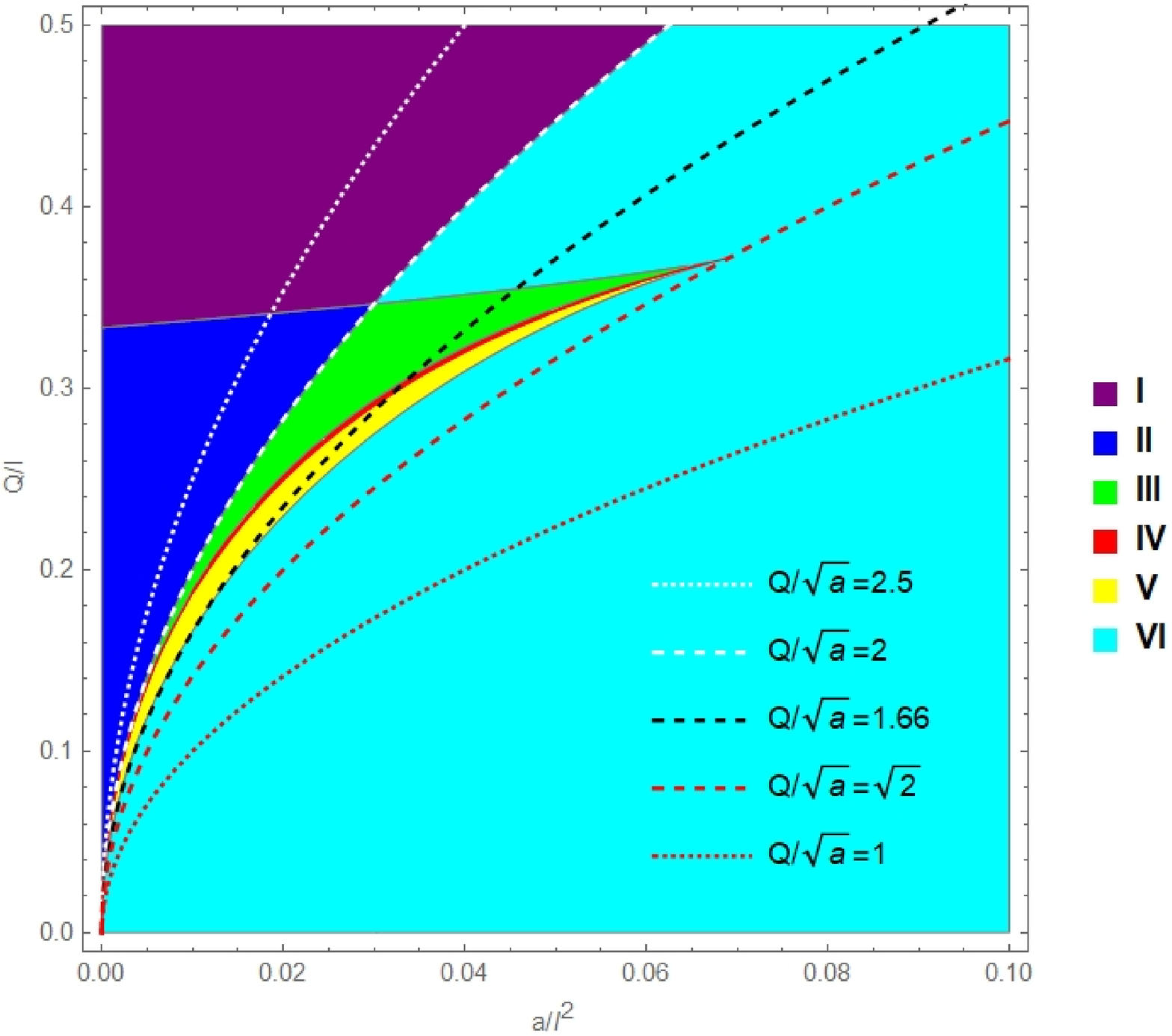}\label{fig:DBIQaP:a}
\includegraphics[width=0.4\textwidth]{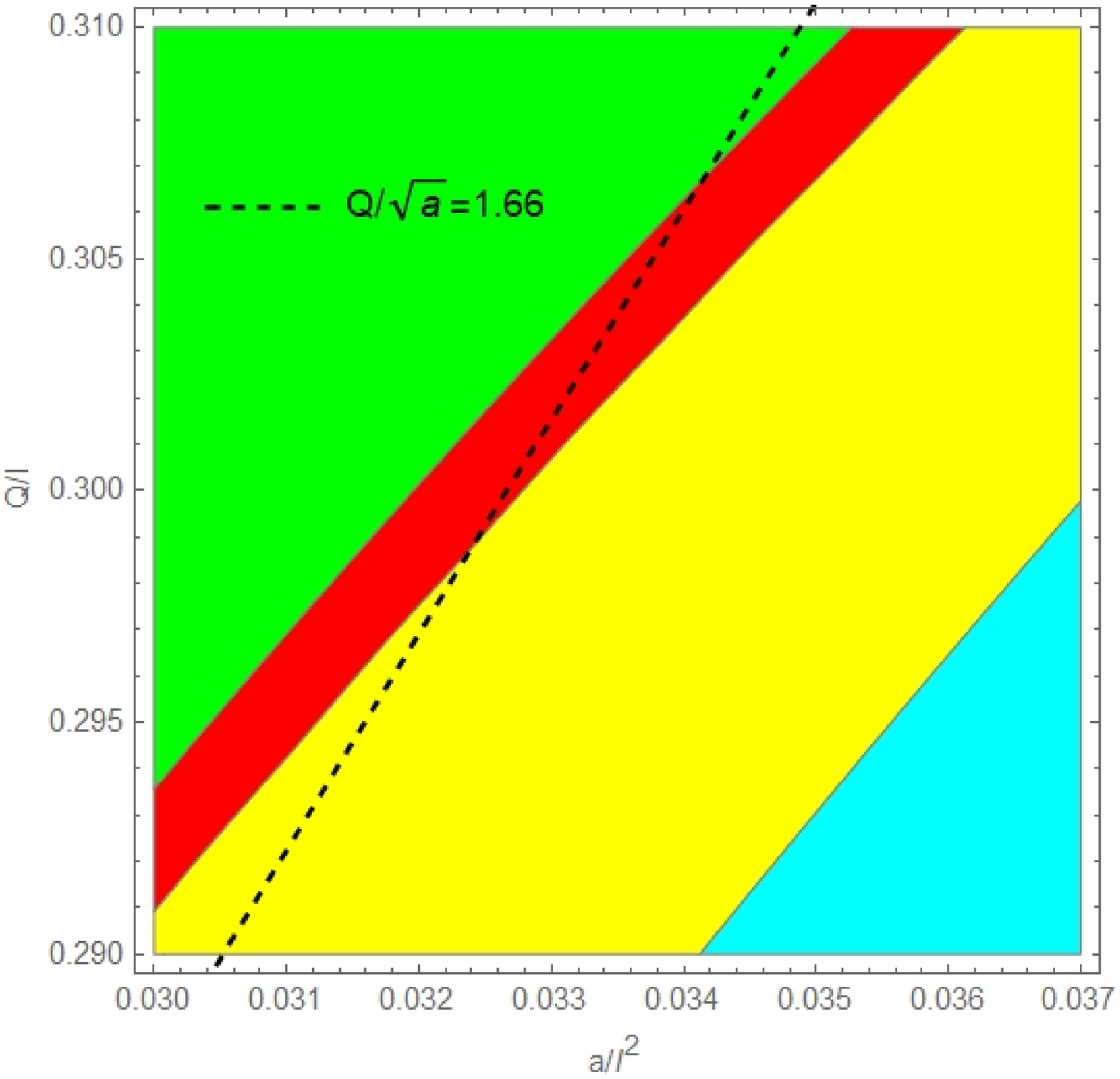}\label{fig:DBIQaP:b}
\end{center}
\caption{{\footnotesize In the case of varying $P$ with fixed $Q$ and $a$, the
system moves along $\tilde{Q}_{l}\left(  \tilde{a}\right)  $, which is
displayed for various values of $Q/\sqrt{a}$. For $Q/\sqrt{a}<\sqrt{2}$, there
is no phase transition in the system. For $Q/\sqrt{a}>\sqrt{2}$, there is one
critical point and the corresponding LBH/SBH first order phase transition. In
addition, for $\sqrt{2}<Q/\sqrt{a}<2$,\ there is a LBH/SBH zeroth order phase
transition occurring in Region IV, corresponding to the LBH/SBH/LBH reentrant
phase transition.}}%
\label{fig:DBIQaP}%
\end{figure}

We now discuss the critical behavior and phase structure of black holes in two
cases. In the first case, $Q$ and $a$ are fixed parameters, and the AdS radius
$l$ (the pressure $P$) varies. With fixed values of $Q$ and $a$, varying $l$
would generate a curve in the $\tilde{a}$-$\tilde{Q}$ plane, which is
determined by%
\begin{equation}
\tilde{Q}_{l}\left(  \tilde{a}\right)  =\frac{Q}{\sqrt{a}}\sqrt{\tilde{a}}.
\end{equation}
In FIG. \ref{fig:DBIQa:b}, we plot $\tilde{Q}_{l}\left(  \tilde{a}\right)  $
for various values of $Q/\sqrt{a}$. It shows that, for $Q/\sqrt{a}<\sqrt{2}$,
there is no critical point for black holes. For $Q/\sqrt{a}>\sqrt{2}$, there
exists one physical critical point. Moreover, the critical behavior is
reminiscent of RN-AdS black holes for $Q/\sqrt{a}>2$. Note that $\tilde{Q}%
_{l}\left(  \tilde{a}\right)  $ intersects the unphysical branch $\tilde
{Q}_{c2}\left(  \tilde{a}\right)  $ for $1.6948>Q/\sqrt{a}>\sqrt{2}$. The
phase structure of $\tilde{Q}_{l}\left(  \tilde{a}\right)  $ can be read from
FIG. \ref{fig:DBIQaP}. It shows that for $Q/\sqrt{a}<2$, $\tilde{Q}_{l}\left(
\tilde{a}\right)  $ is always in Region VI, and hence there is no first order
phase transition. For $Q/\sqrt{a}>2$, as one starts from $P=0$, $\tilde{Q}%
_{l}\left(  \tilde{a}\right)  $ is in Region II, in which there is a first
order phase transition between small BH and large BH. Further increasing $P$,
$\tilde{Q}_{l}\left(  \tilde{a}\right)  $ goes through the critical line and
enters the Region I, in which there is no phase transition. This behavior is
reminiscent of that of the RN-AdS black hole. For $\sqrt{2}<Q/\sqrt{a}<2$, as
$P$ increases from $P=0$, $\tilde{Q}_{l}\left(  \tilde{a}\right)  $ starts
from Region VI, crosses the unphysical critical line and enters Region V,
during which no phase transition occurs. Further increasing $P$, $\tilde
{Q}_{l}\left(  \tilde{a}\right)  $ enters Region IV, in which there is a
reentrant phase transition occurring for some range of $P$. As $P$
continuously increases, $\tilde{Q}_{l}\left(  \tilde{a}\right)  $ enters
Region III, in which a first order phase transition occurs, crosses the
critical line and returns to Region V. The critical behavior and phase
structure in this case has been discussed in \cite{IN-Gunasekaran:2012dq}%
$^{\left[  \ref{ft:1}\right]  }$\footnotetext[1]{\label{ft:1} In
\cite{IN-Gunasekaran:2012dq}, their $b$ is our $\frac{1}{4\sqrt{a}}$.}, which
are correctly reproduced here.

In the second case, $a$ and $P$ $\left(  l\right)  $ are fixed parameters, and
one varies $Q$. FIGs. \ref{fig:DBIQa} show that for $a/l^{2}>\tilde{a}_{c}$,
there is no critical point, and no phase transition occurs. For $a/l^{2}%
<\tilde{a}_{c}$, there is one critical point. As one increases $Q$ from $Q=0$,
the black hole would experience different regions, in which there occur no
phase transition $\rightarrow$ the LBH/SBH/LBH reentrant phase transition
$\rightarrow$ the LBH/SBH first order phase transition $\rightarrow$ no phase
transition. For $a/l^{2}<\tilde{a}_{1}$ and large enough values of $Q$, the
black hole is in Regions I and II, in which the phase transition behavior is
reminiscent of the RN-AdS black hole. The critical behavior and phase
structure in this case has also been studied in \cite{IN-Dehyadegari:2017hvd}.

\begin{figure}[tb]
\begin{center}
\includegraphics[width=0.38\textwidth]{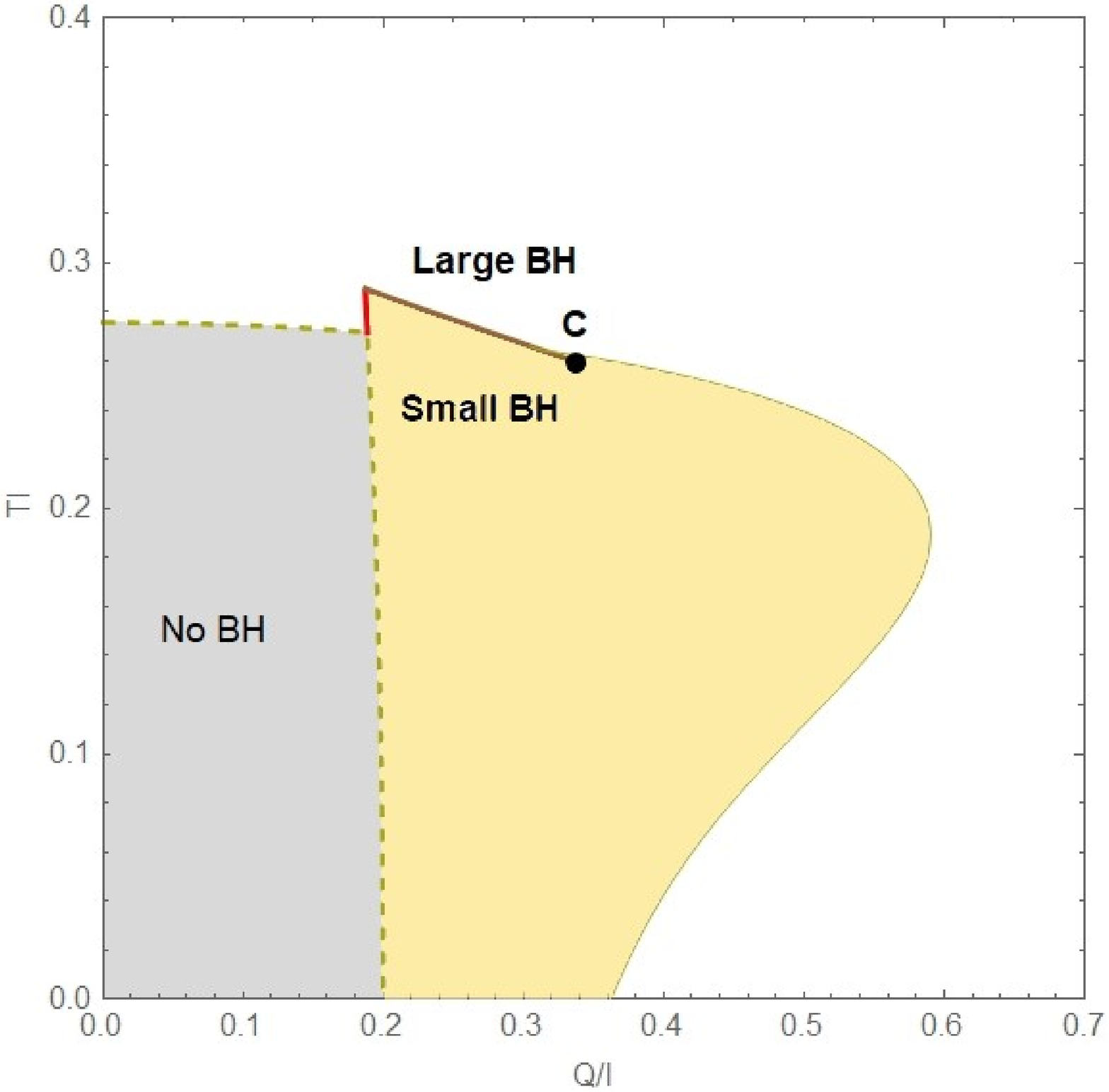}\label{fig:DBIa001:a}
\includegraphics[width=0.395\textwidth]{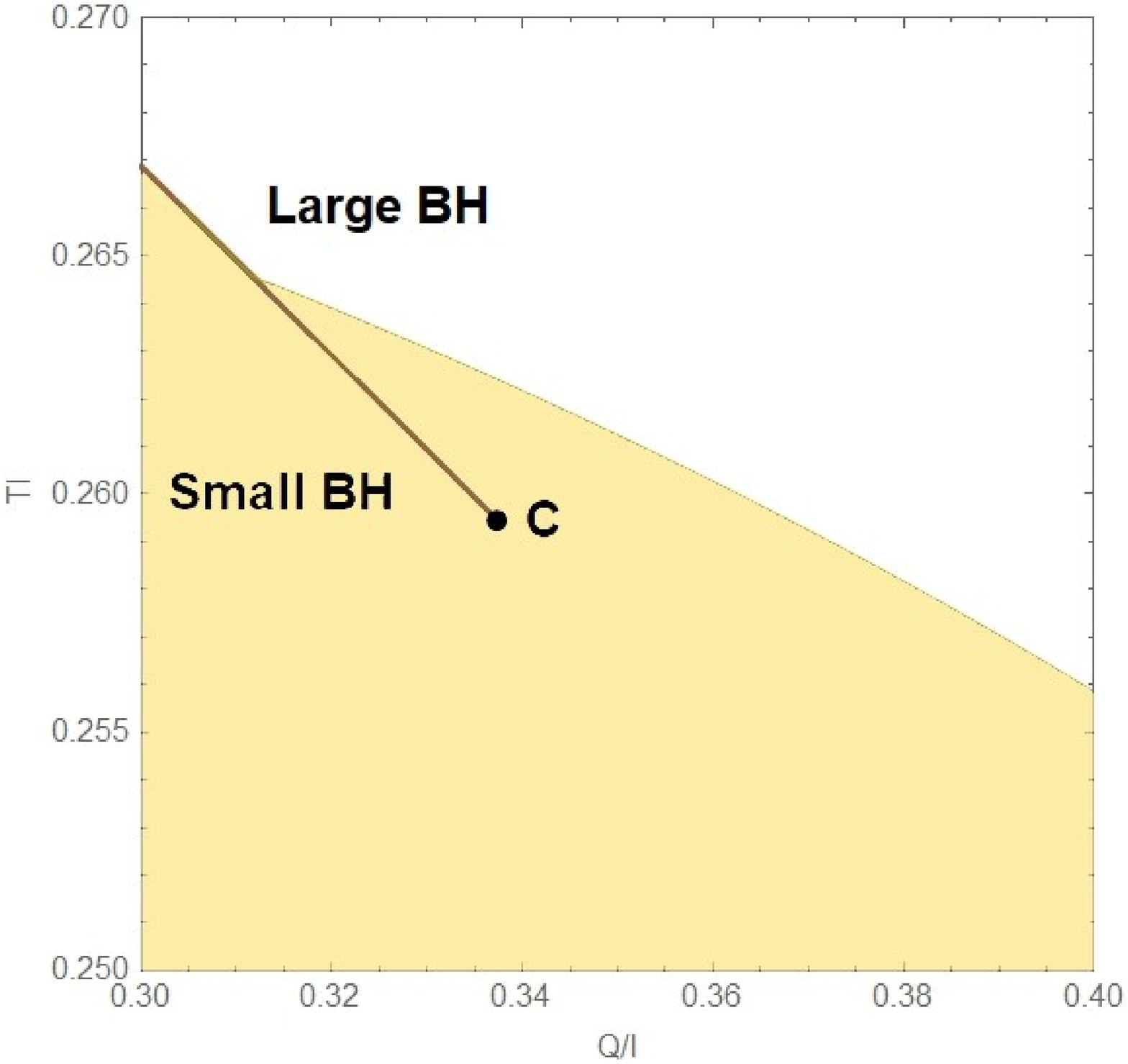}\label{fig:DBIa001:b}
\end{center}
\caption{{\footnotesize The phase diagram in the $\tilde{Q}$-$\tilde{T}$ plane
for BI-AdS black holes with $a/l^{2}=0.01$. The first order phase transition
line separating large BH and small BH is displayed by the brown line, and it
terminates at the critical point, marked by the black dot. There is also a
zeroth order phase transition line, depicted by the red line. All phases in
the diagram are thermally stable. However, the phases in the yellow region are
electrically unstable. Large BH above the first order phase transition line is
always electrically stable except in the region around the critical point,
which are highlighted in the right panel. It shows that the critical point is
in the yellow region.}}%
\label{fig:DBIa001}%
\end{figure}

\begin{figure}[tb]
\begin{center}
\includegraphics[width=0.38\textwidth]{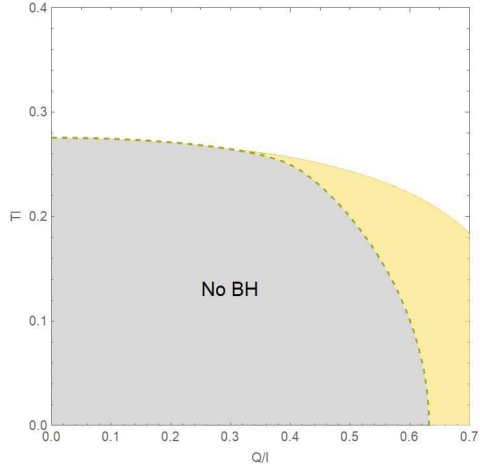}
\end{center}
\caption{{\footnotesize The phase diagram in the $\tilde{Q}$-$\tilde{T}$ plane
for BI-AdS black holes with $a/l^{2}=0.1$. There are no phase transitions. The
black holes in the yellow region are electrically unstable..}}%
\label{fig:DBIa010}%
\end{figure}

The phase diagram of the BI-AdS black hole for $a/l^{2}=0.01$ is displayed in
the $\tilde{Q}$-$\tilde{T}$ plane in FIG. \ref{fig:DBIa001}. There is a
LBH/SBH first order transition for some range of $\tilde{Q}$ and a LBH/SBH
zeroth order phase transition for some smaller range of $\tilde{Q}$. The
zeroth and first order phase transitions are marked by the red and brown
lines, respectively. The first order phase transition line terminates at the
critical point, represented by the black point. No BH region means that no
black hole solutions exist. As discussed before, the black hole solutions in
the phase diagram are thermally stable. However, the solutions in the yellow
region are unstable to electrical fluctuations. Small BH below the first order
phase transition line is always electrically unstable while large BH above the
line is almost electrically stable. The right panel of FIG. \ref{fig:DBIa001}
shows that large BH is only electrically unstable in the region around the
critical point. Note that the black hole solution at the critical point is
electrically unstable.

The phase diagram of the BI-AdS black hole for $a/l^{2}=0.1$ is displayed in
the $\tilde{Q}$-$\tilde{T}$ plane in FIG. \ref{fig:DBIa010}, which is simpler
than that for $a/l^{2}=0.01$. FIG. \ref{fig:DBIQa} shows that when
$a/l^{2}>\tilde{a}_{c}\simeq0.069$ (the black dot), black holes are Regions I
or VI, in which no phase transition occurs. At low temperatures, the black
hole solution is electrically unstable for small enough values of $Q/l$.

\section{iBorn-Infeld AdS Black Hole}

\label{Sec:iBIABH}

We now consider an iBorn-Infeld field with the Lagrangian density%
\begin{equation}
\mathcal{L}\left(  s\right)  =-\frac{1}{a}\left(  1-\sqrt{1+2as}\right)
\text{,}%
\end{equation}
where $a>0$. For an iBorn-Infeld AdS (iBI-AdS) black hole solution, $f\left(
r\right)  $ in the black hole solution $\left(  \ref{eq:ansatz}\right)  $ is
given by%
\begin{equation}
f\left(  r\right)  =1-\frac{M}{8\pi r}+\frac{r^{2}}{l^{2}}-\frac{Q^{2}}%
{6\sqrt{r^{4}-aQ^{2}}+6r^{2}}+\frac{Q^{2}}{3r^{2}}\text{ }_{2}F_{1}\left(
\frac{1}{4},\frac{1}{2},\frac{5}{4};\frac{aQ^{2}}{r^{4}}\right)  ,
\end{equation}
where $M$ and $Q$ are the mass and the charge of the black hole, respectively.
For $A_{t}^{\prime}\left(  r\right)  $, one has
\begin{equation}
A_{t}^{\prime}\left(  r\right)  =\frac{Q}{\sqrt{r^{4}-aQ^{2}}},
\end{equation}
which gives the potential of the black hole%
\begin{equation}
\Phi=\frac{4\pi Q}{r_{+}}\text{ }_{2}F_{1}\left(  \frac{1}{4},\frac{1}%
{2},\frac{5}{4};\frac{aQ^{2}}{r_{+}^{4}}\right)  .
\end{equation}
This iBI-AdS black hole solution has a singularity at $r=r_{s}$, where we
define%
\begin{equation}
r_{s}\equiv a^{1/4}Q^{1/2}.
\end{equation}
To study the nature of this singularity, we compute the corresponding Ricci
scalar:%
\begin{equation}
R=\frac{2}{a}-12+\frac{1}{ar^{2}}\frac{r_{s}^{4}-2r^{4}}{\sqrt{r^{4}-r_{s}%
^{4}}},
\end{equation}
which becomes divergent at $r=r_{s}$. So the singularity at $r=r_{s}$ is a
physical singularity, and one requires that $r>r_{s}$.

The equation of state $\left(  \ref{eq:Ttildal}\right)  $ becomes
\[
\tilde{T}\left(  \tilde{r}_{+}\right)  \equiv\frac{h\left(  \tilde{r}%
_{+}\right)  }{4\pi\tilde{r}_{+}}=\frac{1}{4\pi\tilde{r}_{+}}\left(
1+3\tilde{r}_{+}^{2}-\frac{1}{2}\frac{\tilde{Q}^{2}}{\tilde{r}_{+}^{2}%
+\sqrt{\tilde{r}_{+}^{4}-\tilde{a}\tilde{Q}^{2}}}\right)  ,
\]
where $\tilde{r}_{+}>\tilde{r}_{s}\equiv\tilde{a}^{1/4}\tilde{Q}^{1/2}$,
$\tilde{a}=a/l^{2}$ and $\tilde{Q}=Q/l$. It can show that $h\left(  \tilde
{r}_{+}\right)  $ is a strictly increasing function and $h\left(
\infty\right)  \rightarrow+\infty$. At $\tilde{r}_{+}=\tilde{r}_{s}$, one has%
\begin{equation}
h\left(  \tilde{r}_{s}\right)  =1+3\tilde{a}^{1/2}\tilde{Q}-\frac{\tilde{Q}%
}{2\tilde{a}^{1/2}}.
\end{equation}
For $h(\tilde{r}_{s})\leq0$, which reduces to%
\begin{equation}
\tilde{a}<\frac{1}{6}\text{ and }\tilde{Q}\geq\frac{2\sqrt{\tilde{a}}%
}{1-6\tilde{a}},
\end{equation}
$\tilde{T}\left(  \tilde{r}_{+}\right)  =0$ has one solution $\tilde{r}%
_{+}=\tilde{r}_{e}$, at which the black hole becomes extremal. In this case,
the black hole is RN type. For $h\left(  \tilde{r}_{s}\right)  >0$, which
reduces to%
\begin{equation}
\tilde{a}<\frac{1}{6}\text{ and }\tilde{Q}<\frac{2\sqrt{\tilde{a}}}%
{1-6\tilde{a}}\text{; or }\tilde{a}\geq\frac{1}{6}\text{, }%
\end{equation}
the temperature of the black hole has a positive minimum value, and the black
hole is Schwarzschild-like type.

The equation $\tilde{T}^{\prime\prime}\left(  \tilde{r}_{+}\right)  =0$
becomes
\begin{equation}
z\left(  x\right)  \equiv x^{3}-\frac{3\tilde{Q}^{2}}{2}x^{2}-\tilde{a}%
\tilde{Q}^{4}=0,
\end{equation}
where $x=\sqrt{\tilde{r}_{+}^{4}-\tilde{a}\tilde{Q}^{2}}>0$. It can show that
$z\left(  x\right)  $ has a local maximum of $z\left(  0\right)  =-\tilde
{a}\tilde{Q}^{4}<0$ at $x=0$ and a local minimum of $z(\tilde{Q}^{2}%
)=(-\tilde{Q}^{2}/2-\tilde{a})\tilde{Q}^{4}<0$ at $x=\tilde{Q}^{2}$. So
$z\left(  x\right)  =0$ always admits one single positive real root
$x=x_{1}>0$. Since $\lim_{\tilde{r}_{+}\rightarrow\tilde{r}_{s}}\tilde
{T}^{\prime}\left(  \tilde{r}_{+}\right)  =+\infty$ and $\lim_{\tilde{r}%
_{+}\rightarrow+\infty}\tilde{T}^{\prime}\left(  \tilde{r}_{+}\right)
=3\sqrt{\tilde{a}}\tilde{Q}$, $\tilde{T}^{\prime}\left(  \tilde{r}_{+}\right)
$ always has a global minimum of $\tilde{T}_{\text{min}}^{\prime}\equiv
\tilde{T}^{\prime}\left(  \tilde{r}_{1}\right)  $ at $\tilde{r}_{+}=\tilde
{r}_{1}\equiv(x_{1}^{2}+\tilde{a}\tilde{Q}^{2})^{1/4}$. In what follows, we
also find six regions in the $\tilde{a}$-$\tilde{Q}$ plane for iBI-AdS black
holes, in each of which the black hole has the distinct behavior of the
branches and the phase structure:

\begin{figure}[tbh]
\begin{center}
\subfigure[{~\scriptsize Region I: $a/l^{2}=0.01$ and $Q/l=0.5$. There is no phase transition.}]{
\includegraphics[width=1.01\textwidth]{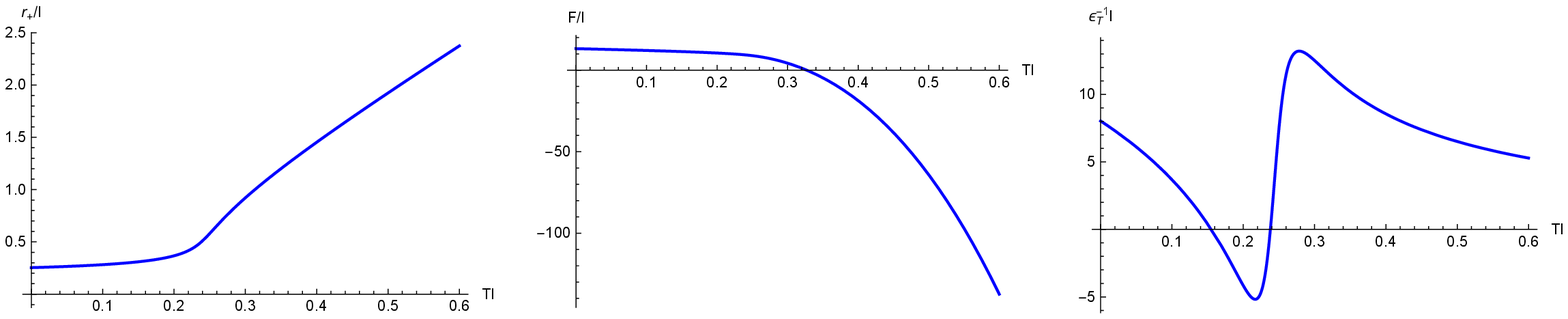}\label{fig:iDBIR12:a}}
\subfigure[{~\scriptsize Region II: $a/l^{2}=0.01$ and $Q/l=0.22$. There is a first order phase transition between small BH and large BH.}]{
\includegraphics[width=1.01\textwidth]{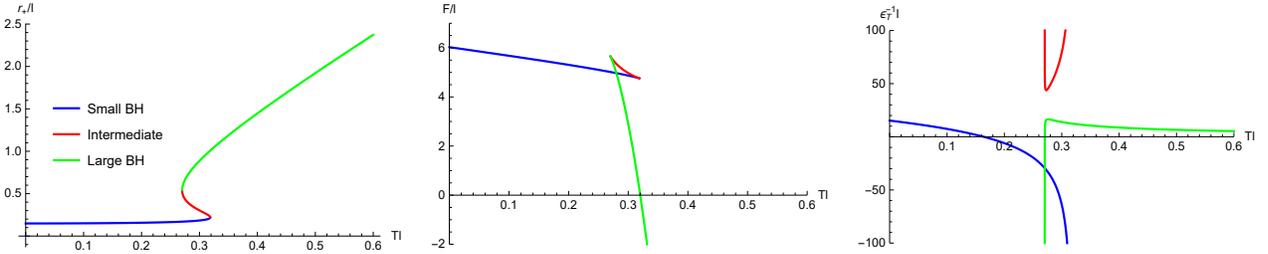}\label{fig:iDBIR12:b}}
\end{center}
\caption{{\footnotesize Plot of $\tilde{r}_{+}$, $\tilde{F}$ and $\epsilon
_{T}^{-1}l$ against $\tilde{T}$ for iBI-AdS black holes in Regions I and II,
where Black holes in these regions are RN type. Regions I and II can be
considered as reminiscent of RN-AdS black holes. The blue and green branches
are thermally stable. It shows that for small enough $\tilde{a}$, e.g.,
$\tilde{a}=0.01$, the small BH branch is electrically stable for $\tilde
{T}<\tilde{T}_{1}$ and unstable for $\tilde{T}>\tilde{T}_{1}$ with some
$\tilde{T}_{1}>0$.}}%
\label{fig:iDBIR12}%
\end{figure}

\begin{figure}[ptb]
\begin{center}
\subfigure[{~\scriptsize Region III: $a/l^{2}=0.05$ and $Q/l=0.5$. There is no phase transition.}]{
\includegraphics[width=1\textwidth]{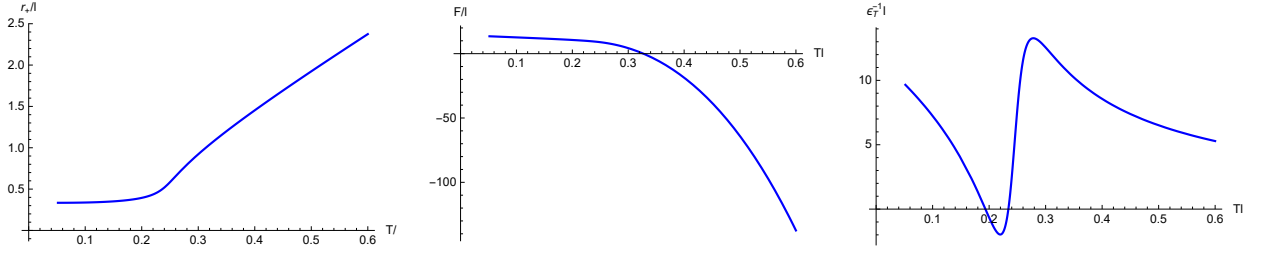}\label{fig:iDBIR36:a}}
\subfigure[{~ \scriptsize Region IV: $a/l^{2}=0.05$ and $Q/l=0.23$. There is a first order phase transition between small BH and large BH.}]{
\includegraphics[width=1\textwidth]{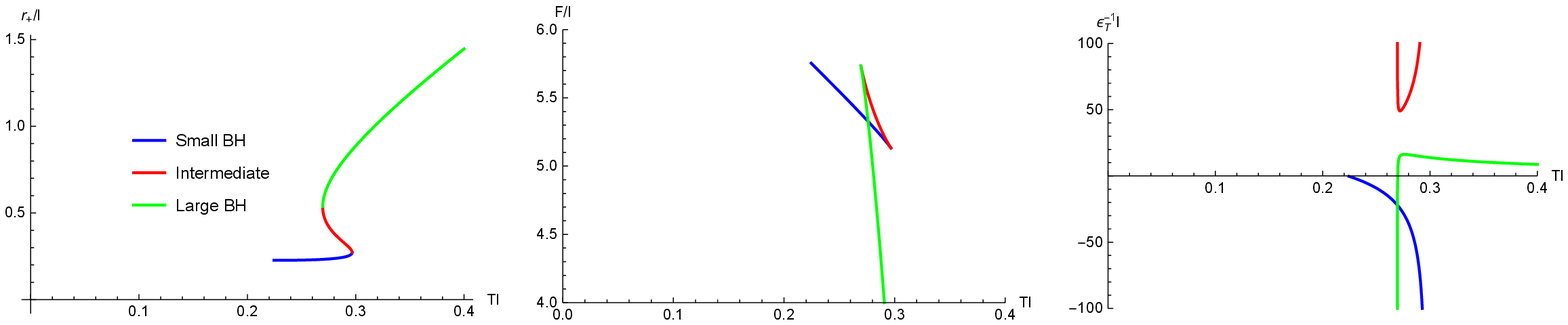}\label{fig:iDBIR36:b}}
\subfigure[{~ \scriptsize Region V: $a/l^{2}=0.05$ and $Q/l=0.184$. The arrows in the inset
indicate increasing $\tilde{T}$. As $\tilde{T}$ increases, the black hole
jumps from the large BH branch to the small BH one, corresponding to the
zeroth order phase transition between small BH and large BH. Further
increasing $\tilde{T}$, there would be a first order phase transition
returning to large BH. Here we observe
LBH/SBH/LBH reentrant phase transition. }]{
\includegraphics[width=1\textwidth]{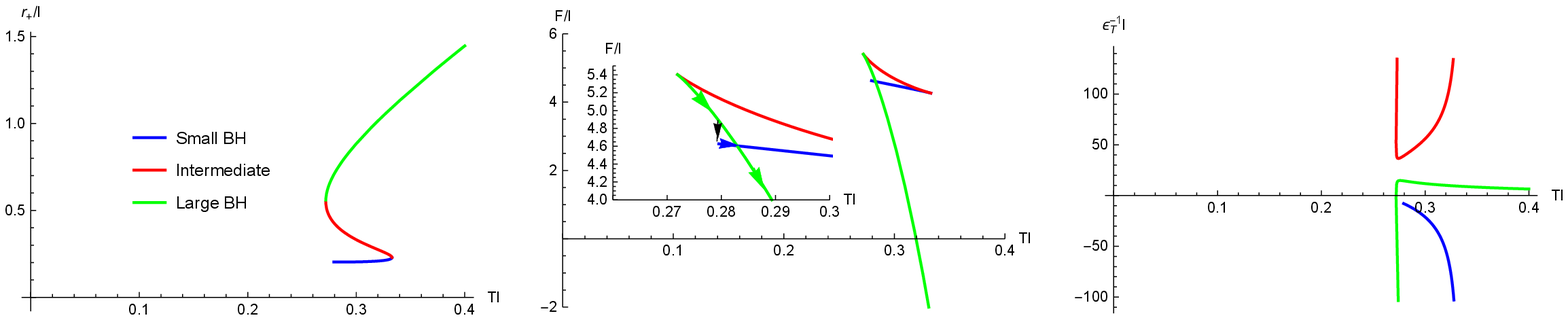}\label{fig:iDBIR36:c}}
\subfigure[{~ \scriptsize Region VI: $a/l^{2}=0.05$ and $Q/l=0.17$. There is no phase transition.}]{
\includegraphics[width=1\textwidth]{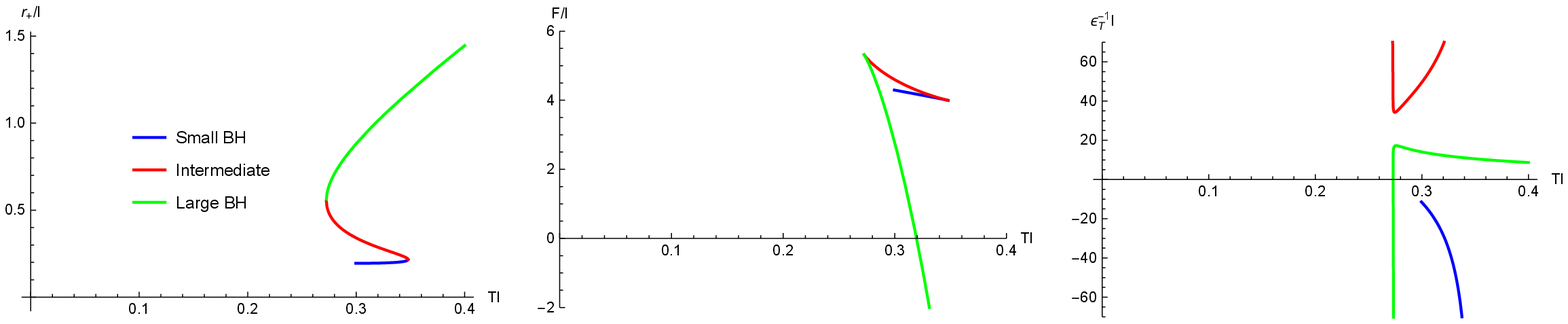}\label{fig:iDBIR36:d}}
\end{center}
\caption{{\footnotesize Plot of $\tilde{r}_{+}$, $\tilde{F}$ and $\epsilon
_{T}^{-1}l$ against $\tilde{T}$ for iBI-AdS black holes in Regions III, IV,V
and VI. The temperature of black holes in these regions has a minimum value
greater than zero. The blue and green branches are always thermally stable. It
shows that for large enough $\tilde{a}$, e.g., $\tilde{a}=0.05$, the small BH
branch is always electrically unstable.}}%
\label{fig:iDBIR36}%
\end{figure}

\begin{itemize}
\item Region I: $h\left(  \tilde{r}_{s}\right)  \leq0$ and $\tilde
{T}_{\text{min}}^{\prime}\geq0$. Since $\tilde{T}^{\prime}\left(  \tilde
{r}_{+}\right)  \geq\tilde{T}_{\text{min}}^{\prime}\geq0$, $\tilde{T}\left(
\tilde{r}_{+}\right)  $ is\ an increasing function in this region. So there is
only one thermally stable branch for $\tilde{r}_{+}(\tilde{T})$. We plot the
radius $\tilde{r}_{+}$, the Gibbs energy $\tilde{F}$ and the isothermal
permittivity $\epsilon_{T}^{-1}l$ as functions of $\tilde{T}$ in FIG.
\ref{fig:iDBIR12:a} for a black hole with $\tilde{a}=0.01$ and $\tilde{Q}=0.5$
in this region. Moreover, this black hole is electrically stable for small
enough and large enough values of $\tilde{T}$. However for large enough
$\tilde{a}$, the black hole is always electrically stable.

\item Region II: $h\left(  \tilde{r}_{s}\right)  \leq0$ and $\tilde
{T}_{\text{min}}^{\prime}<0$. In this region, $\tilde{T}^{\prime}\left(
\tilde{r}_{+}\right)  =0$ has two solutions $\tilde{r}_{+}=$\ $\tilde
{r}_{\text{max}}$ and\ $\tilde{r}_{\text{min}}$ with $\tilde{r}_{\text{max}%
}<\tilde{r}_{1}<\tilde{r}_{\text{min}}$. Since $\tilde{T}\left(
+\infty\right)  =+\infty$, $\tilde{T}\left(  \tilde{r}_{+}\right)  $ has a
local maximum of $\tilde{T}_{\text{max}}\equiv\tilde{T}\left(  \tilde
{r}_{\text{max}}\right)  $ at $\tilde{r}_{+}=$\ $\tilde{r}_{\text{max}}$ and a
local minimum of $\tilde{T}_{\text{min}}\equiv\tilde{T}\left(  \tilde
{r}_{\text{min}}\right)  $ at $\tilde{r}_{+}=$\ $\tilde{r}_{\text{min}}$.
There are three branches for $\tilde{r}_{+}(\tilde{T})$: small BH for
$0\leq\tilde{T}\leq\tilde{T}_{\text{max}}$, intermediate BH for $\tilde
{T}_{\text{min}}\leq\tilde{T}\leq\tilde{T}_{\text{max}}$ and large BH for
$\tilde{T}\geq\tilde{T}_{\text{min}}$, which are displayed in the left panel
of FIG. \ref{fig:iDBIR12:b}. The middle panel shows that there is a first
order phase transition between small BH and large BH occurring at $\tilde
{T}_{\text{min}}\leq\tilde{T}\leq\tilde{T}_{\text{max}}$. Both the small BH
and large BH branches are thermally stable. The right panel shows that large
BH is almost electrically stable. However, the electrical stability of small
BH depends on the values of $\tilde{a}$. For small enough $\tilde{a}$, e.g.,
$\tilde{a}=0.01$, small BH is electrically stable for small enough $\tilde{T}%
$. For large enough $\tilde{a}$, small BH is always electrically unstable.

\item Region III: $h\left(  \tilde{r}_{s}\right)  >0$ and $\tilde
{T}_{\text{min}}^{\prime}\geq0$. As shown in FIG. \ref{fig:iDBIR36:a}, the
black hole's temperature has a minimum of $\frac{h\left(  \tilde{r}%
_{s}\right)  }{4\pi r_{s}}$ at $\tilde{r}_{+}=\tilde{r}_{s}$. There is only
one branch for $\tilde{r}_{+}(\tilde{T})$ in this region, which is thermally
stable. Similarly to Region I, the black hole is electrically unstable for
some finite range of $\tilde{T}$ for small enough $\tilde{a}$. However for
large enough $\tilde{a}$, the black hole is always electrically stable.

\item Region IV: $h\left(  \tilde{r}_{s}\right)  >0$, $\tilde{T}_{\text{min}%
}^{\prime}<0$ and $\tilde{T}_{\text{min}}>\tilde{T}\left(  \tilde{r}%
_{s}\right)  $. In this region, $\tilde{T}^{\prime}\left(  \tilde{r}%
_{+}\right)  =0$ has two solutions $\tilde{r}_{+}=$\ $\tilde{r}_{\text{max}}$
and\ $\tilde{r}_{+}=\tilde{r}_{\text{min}}$ with $\tilde{r}_{s}<\tilde
{r}_{\text{max}}<\tilde{r}_{1}<\tilde{r}_{\text{min}}$. So $\tilde{T}\left(
\tilde{r}_{+}\right)  $ has a local maximum of $\tilde{T}_{\text{max}}$ at
$\tilde{r}_{+}=$\ $\tilde{r}_{\text{max}}$, a local minimum of $\tilde
{T}_{\text{min}}$ at $\tilde{r}_{+}=$\ $\tilde{r}_{\text{min}}$ and a global
minimum of $\tilde{T}\left(  \tilde{r}_{s}\right)  $ at $\tilde{r}_{+}%
=$\ $\tilde{r}_{s}$. There are three branches for $\tilde{r}_{+}(\tilde{T})$:
small BH for $\tilde{T}\left(  \tilde{r}_{s}\right)  \leq\tilde{T}\leq
\tilde{T}_{\text{max}}$, intermediate BH for $\tilde{T}_{\text{min}}\leq
\tilde{T}\leq\tilde{T}_{\text{max}}$ and large BH for $\tilde{T}\geq\tilde
{T}_{\text{min}}$, which are displayed in the left panel of FIG.
\ref{fig:iDBIR36:b}. There is a first order phase transition between small BH
and large BH occurring at $\tilde{T}_{\text{min}}\leq\tilde{T}\leq\tilde
{T}_{\text{max}}$. Both the small BH and large BH branches are thermally
stable, while the intermediate BH branch is not. The electrical stability of
black holes in this region is similar to that in Region II.

\item Region V: $h\left(  \tilde{r}_{s}\right)  >0$, $\tilde{T}_{\text{min}%
}^{\prime}<0$, $\tilde{T}_{\text{min}}\leq\tilde{T}\left(  \tilde{r}%
_{s}\right)  $ and $\tilde{F}_{S}\left(  \tilde{r}_{s}\right)  <\tilde{F}%
_{L}\left(  \tilde{r}_{s}\right)  $, where $\tilde{F}_{S/L}$ is the Gibbs free
energy of the small/large BH branch. In this region, $\tilde{T}\left(
\tilde{r}_{+}\right)  $ has a local maximum of $\tilde{T}_{\text{max}}$ at
$\tilde{r}_{+}=$\ $\tilde{r}_{\text{max}}$, a global minimum of $\tilde
{T}_{\text{min}}$ at $\tilde{r}_{+}=$\ $\tilde{r}_{\text{min}}$ and a local
minimum of $\tilde{T}\left(  \tilde{r}_{s}\right)  $ at $\tilde{r}_{+}%
=$\ $\tilde{r}_{s}$. FIG. \ref{fig:iDBIR36:c}\ shows that there are three
branches of $\tilde{r}_{+}(\tilde{T})$ in this region.\ The Gibbs free energy
of the three branches is plotted in\ the middle panel. As $\tilde{T}$
increases from $\tilde{T}_{\text{min}}$, the black hole follows direction of
arrows in the inset. It shows that\ at $\tilde{T}=\tilde{T}\left(  \tilde
{r}_{s}\right)  $, there is a finite jump in Gibbs free energy leading to a
zeroth order phase transition from large BH to small BH. Further increasing
$\tilde{T}$, a first order phase transition returning to large BH occurs at
$\tilde{T}\left(  \tilde{r}_{s}\right)  \leq\tilde{T}\leq\tilde{T}%
_{\text{max}}$. This LBH/SBH/LBH transition corresponds to a reentrant phase transition.

\item Region VI: $h\left(  \tilde{r}_{s}\right)  >0$, $\tilde{T}_{\text{min}%
}^{\prime}<0$, $\tilde{T}_{\text{min}}\leq\tilde{T}\left(  \tilde{r}%
_{s}\right)  $ and $\tilde{F}_{S}\left(  \tilde{r}_{s}\right)  \geq\tilde
{F}_{L}\left(  \tilde{r}_{s}\right)  $, As shown in the left panel of FIG.
\ref{fig:iDBIR36:d}, there are three branches of $\tilde{r}_{+}(\tilde{T})$ in
this region. The middle panel shows that the large BH branch is always
thermodynamically preferred for $\tilde{T}\geq\tilde{T}_{\text{min}}$, and
hence there is no phase transition in this region.
\end{itemize}

\begin{figure}[tb]
\begin{center}
\includegraphics[width=0.5\textwidth]{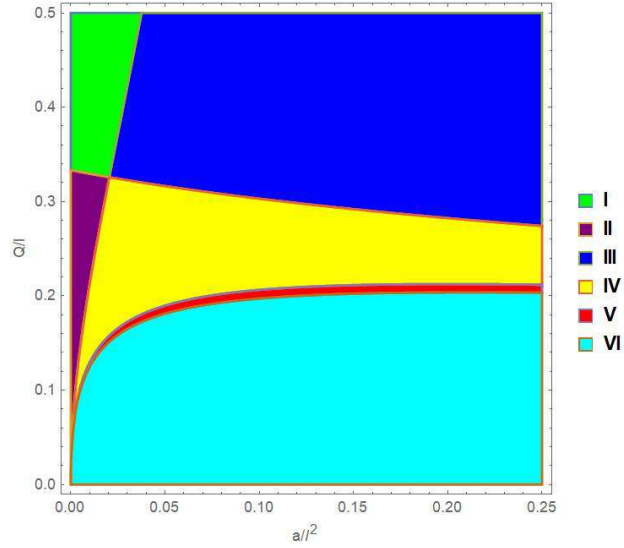}
\end{center}
\caption{{\footnotesize The six regions in the $\tilde{a}$-$\tilde{Q}$ plane,
each of which possesses the distinct behavior of the branches and the phase
structure for iBI-AdS black holes. The LBH/SBH/LBH reentrant phase transition
occurs in Region V. The LBH/SBH first order phase transition occurs in Regions
II and IV. No phase transitions occur in Regions I,III and VI.}}%
\label{fig:iDBIQa}%
\end{figure}

\begin{figure}[tb]
\begin{center}
\subfigure[{~ \scriptsize There is always one physical critical point for the black holes on
$\tilde{Q}_{l}\left(  \tilde{a}\right)  $. For $Q/\sqrt{a}>\tilde{Q}_{1}\simeq2.28$, the inset displays that the critical point occurs for the
Schwarzschild-like type black hole.}]{
\includegraphics[width=0.38\textwidth]{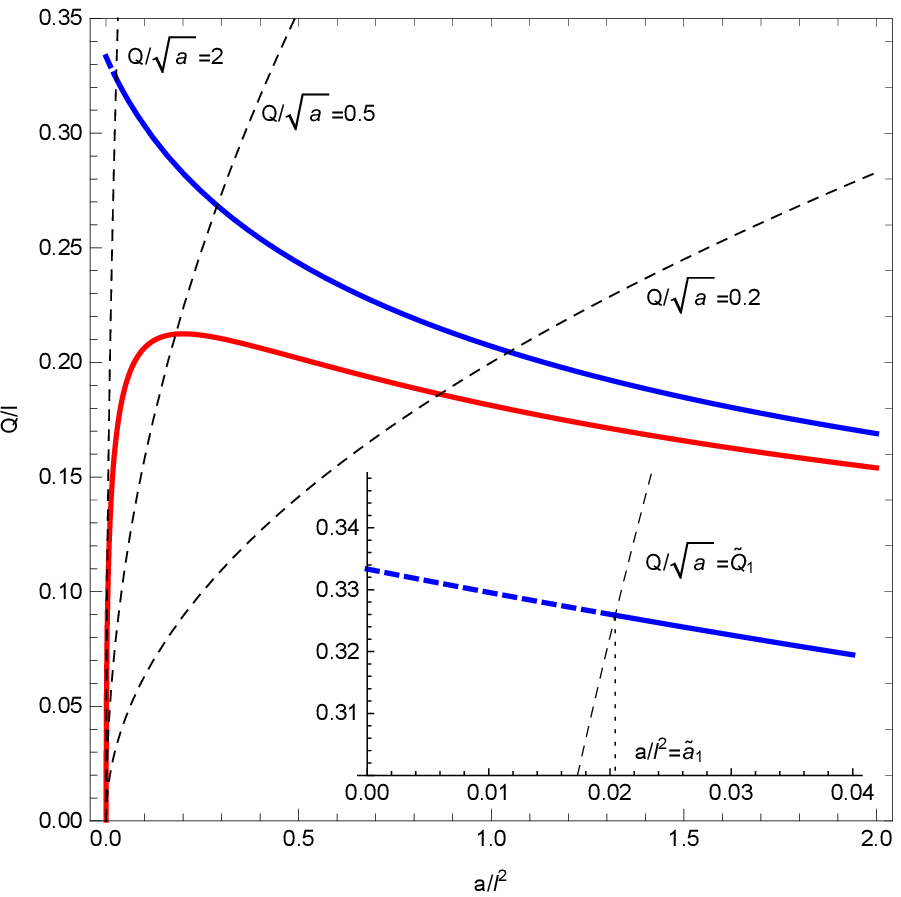}\label{fig:iDBIc:a}}
\subfigure[{~ \scriptsize For $Q/\sqrt{a}<2$, $\tilde{Q}_{l}\left(  \tilde{a}\right)  $ always intersect
with $\tilde{Q}_{45}\left(  \tilde{a}\right)  $. In this case, the black holes
on $\tilde{Q}_{l}\left(  \tilde{a}\right)  $ could be in Region V for some
range of $P$, where the reentrant phase transition occurs.}]{
\includegraphics[width=0.38\textwidth]{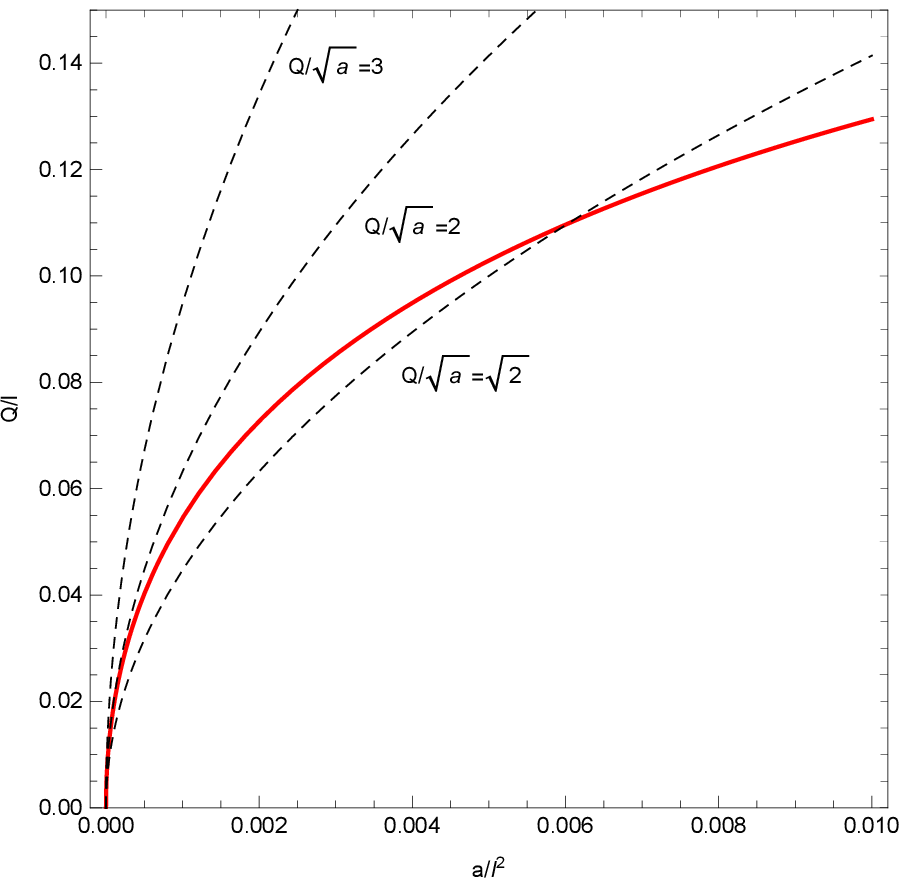}\label{fig:iDBIc:b}}
\end{center}
\caption{{\footnotesize The blue line is the critical line, which consists of
the boundaries $\tilde{Q}_{12}\left(  \tilde{a}\right)  $ (the blue dashed
line) and $\tilde{Q}_{34}\left(  \tilde{a}\right)  $ (the blue solid line).
The black hole on $\tilde{Q}_{12}\left(  \tilde{a}\right)  $ is RN type. The
red line is the boundary $\tilde{Q}_{45}\left(  \tilde{a}\right)  $. In the
case of varying $P$ with fixed values of $Q$ and $a$, the system moves along
$\tilde{Q}_{l}\left(  \tilde{a}\right)  $, which is plotted for various values
of $Q/\sqrt{a}$. }}%
\label{fig:iDBIc}%
\end{figure}

These six regions are plotted in the $\tilde{a}$-$\tilde{Q}$ plane in FIG.
\ref{fig:iDBIQa}, from which the critical line can be read. In fact, the
critical line is determined by $\tilde{T}^{\prime}\left(  \tilde{r}%
_{1}\right)  =0$ and hence is composed of $\tilde{Q}_{12}\left(  \tilde
{a}\right)  $ and $\tilde{Q}_{34}\left(  \tilde{a}\right)  $, where $\tilde
{Q}_{ij}\left(  \tilde{a}\right)  $ is the boundary between Region $i$ and
Region $j$. The critical line is plotted in FIG. \ref{fig:iDBIc:a}, and
$\tilde{Q}_{12}\left(  \tilde{a}\right)  /\tilde{Q}_{34}\left(  \tilde
{a}\right)  $ is depicted by the blue dashed/solid line. The inset in FIG.
\ref{fig:iDBIc:a} demonstrates that for $\tilde{a}\leq\tilde{a}_{1}\simeq
0.02$, the critical line is $\tilde{Q}_{12}\left(  \tilde{a}\right)  $, on
which black holes are RN type. Moreover, the critical line of iBI-AdS black
holes in the $\tilde{a}$-$\tilde{Q}$ plane is semi-infinite while that of
BI-AdS black holes is a finite line. According to FIGs. \ref{fig:iDBIR12:b}
and \ref{fig:iDBIR36:b}, the critical line is physical since it globally
minimizes the Gibbs free energy.

In the case of varying $P$ with fixed values of $Q$ and $a$, the system moves
along $\tilde{Q}_{l}\left(  \tilde{a}\right)  =\left(  Q/\sqrt{a}\right)
\sqrt{\tilde{a}}$ in the $\tilde{a}$-$\tilde{Q}$ plane, which are plotted for
various values of $Q/\sqrt{a}$ in FIG. \ref{fig:iDBIc}. FIG. \ref{fig:iDBIc:a}
shows that there always exists one physical critical point. For $Q/\sqrt
{a}\leq\tilde{Q}_{1}\simeq2.28$, the inset in FIG. \ref{fig:iDBIc:a} shows
that the critical point occurs for the RN type black hole. The boundary
$\tilde{Q}_{45}\left(  \tilde{a}\right)  $ is displayed by the red line in
FIG. \ref{fig:iDBIc}. If $\tilde{Q}_{l}\left(  \tilde{a}\right)  $ intersects
with $\tilde{Q}_{45}\left(  \tilde{a}\right)  $, the black holes on $\tilde
{Q}_{l}\left(  \tilde{a}\right)  $ could be in Region V for some range of $P$.
The numerical result and FIG. \ref{fig:iDBIc:b} show that when $Q/\sqrt{a}<2$,
$\tilde{Q}_{l}\left(  \tilde{a}\right)  $ always intersects with $\tilde
{Q}_{45}\left(  \tilde{a}\right)  $, and there is a reentrant phase transition
occurring for some range of $P$. Thus for $Q/\sqrt{a}<2$, as $P$ continuously
increases from $P=0$, the black holes on $\tilde{Q}_{l}\left(  \tilde
{a}\right)  $ experience the following regions: Region VI (no phase
transitions) $\rightarrow$ Region V (the LBH/SBH/LBH reentrant phase
transition) $\rightarrow$ Regions II or IV (the LBH/SBH first order phase
transition) $\rightarrow$ Regions I or III (no phase transitions). For
$Q/\sqrt{a}>2$, there is a critical point on $\tilde{Q}_{l}\left(  \tilde
{a}\right)  $ occurring at $P=P_{c}$. For $P<P_{c}$, the black holes on
$\tilde{Q}_{l}\left(  \tilde{a}\right)  $ with $Q/\sqrt{a}>2$ are in Regions
II or IV, and there is a first order phase transition between small BH and
large BH. For $P>P_{c}$, they are in Regions I or III, and no phase transition occurs.

In the case of varying $Q$ with fixed values of $P$ and $a$, the system moves
along a constant-$\tilde{a}$ line in the $\tilde{a}$-$\tilde{Q}$ plane. FIG.
\ref{fig:iDBIc:a} shows that constant-$\tilde{a}$ lines always intersect the
critical line and the boundary $\tilde{Q}_{45}\left(  \tilde{a}\right)  $. As
one increases $Q$ from $Q=0$, the black holes on a constant-$\tilde{a}$ line
experience the following regions: Region VI (no phase transitions)
$\rightarrow$ Region V (the LBH/SBH/LBH reentrant phase transition)
$\rightarrow$ Regions II or IV (the LBH/SBH first order phase transition)
$\rightarrow$ Regions I or III (no phase transitions). For $\tilde{a}\geq1/6$,
the black holes on a constant-$\tilde{a}$ line are always Schwarzschild-like
type.\begin{figure}[tb]
\begin{center}
\subfigure[{~ \scriptsize The phase diagram for $a/l^{2}=0.01$.}]{
\includegraphics[width=0.37\textwidth]{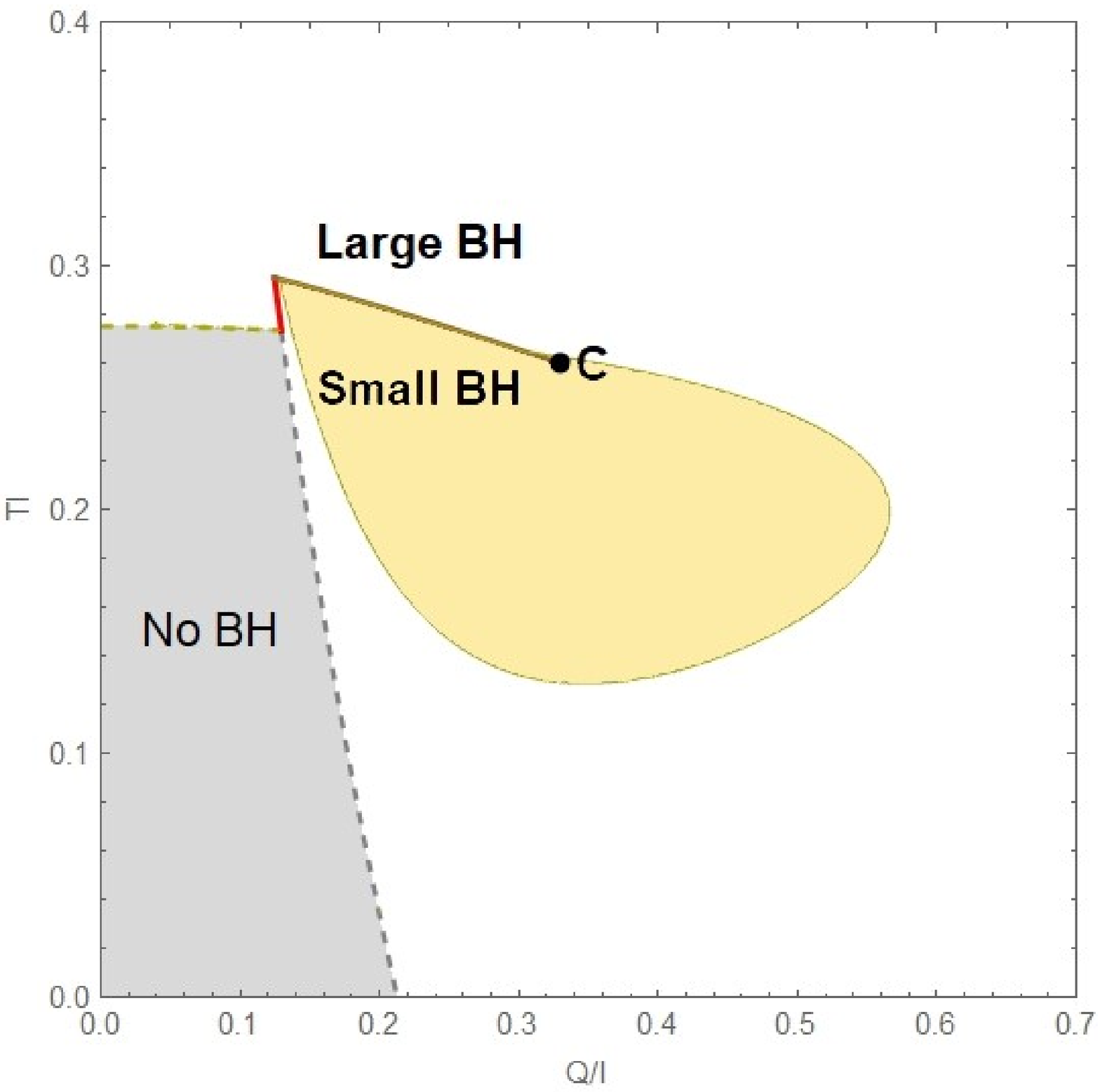}\label{fig:iDBIQT:a}}
\subfigure[{~ \scriptsize Hightlighted region of the reentrant phase transition in the phase diagram for
$a/l^{2}=0.01$.}]{
\includegraphics[width=0.385\textwidth]{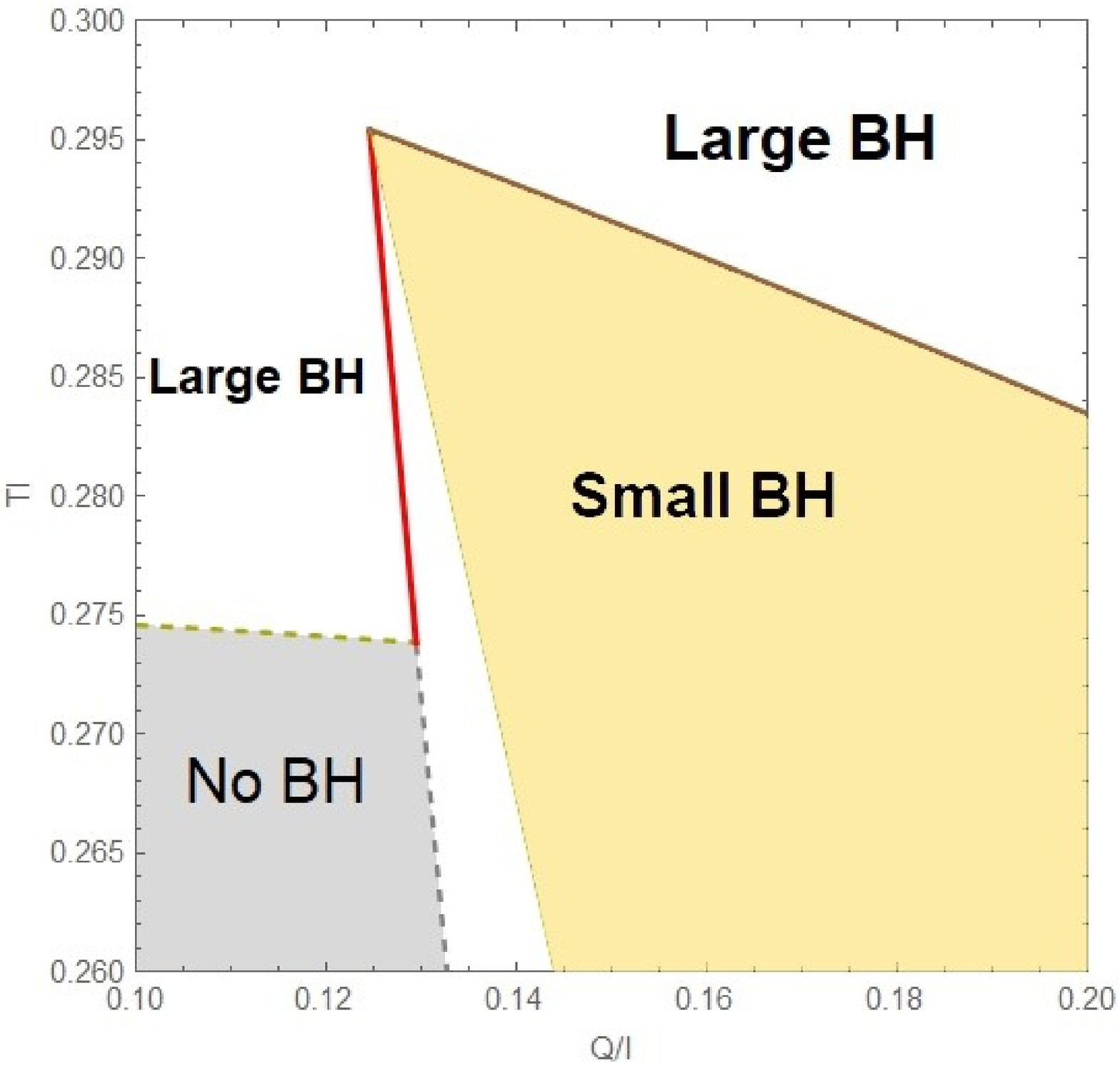}\label{fig:iDBIQT:b}}
\subfigure[{~ \scriptsize Hightlighted region near the critical point in the phase diagram for
$a/l^{2}=0.01$.}]{
\includegraphics[width=0.385\textwidth]{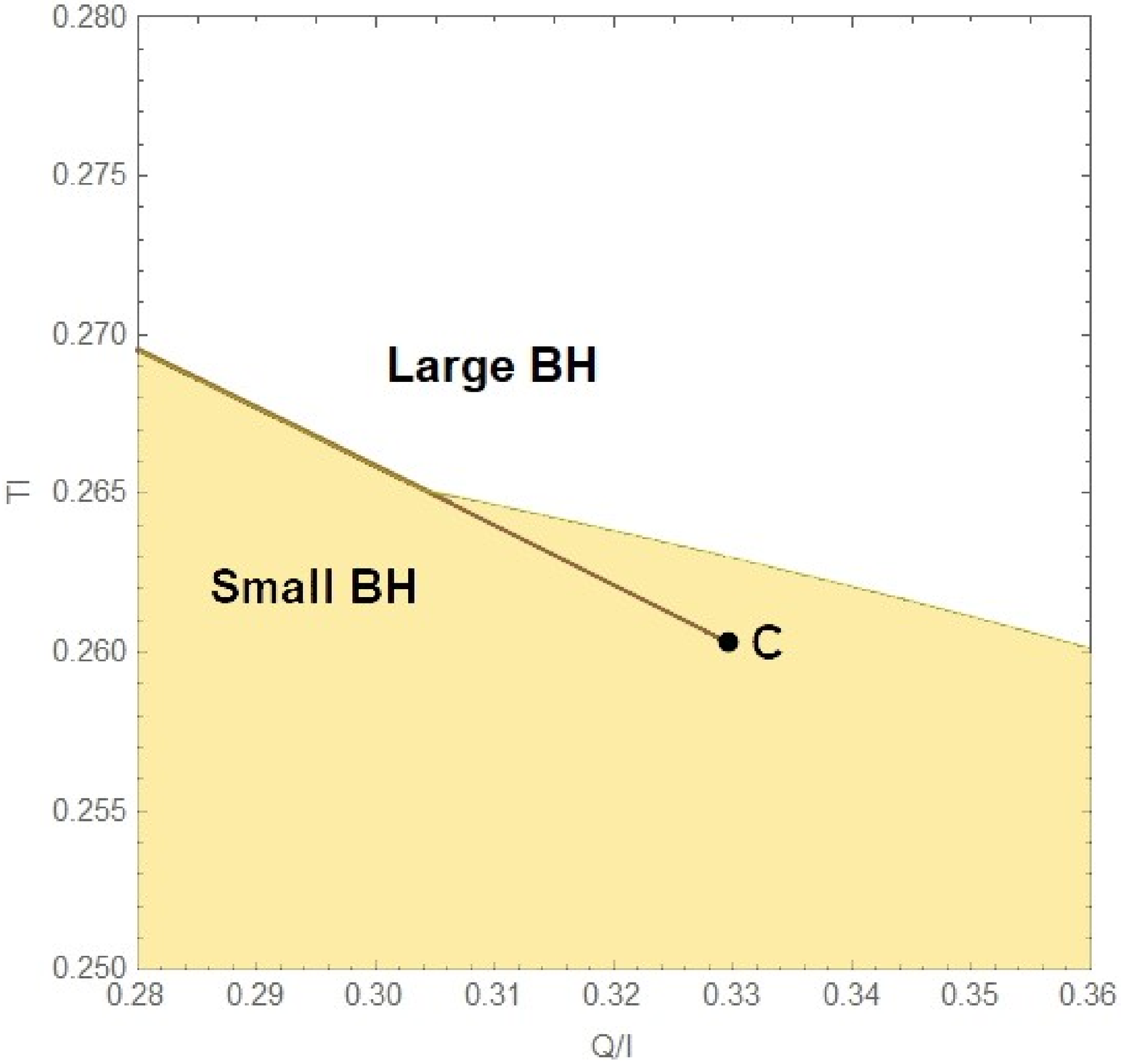}\label{fig:iDBIQT:c}}
\subfigure[{~ \scriptsize The phase diagram for $a/l^{2}=0.1$.}]{
\includegraphics[width=0.37\textwidth]{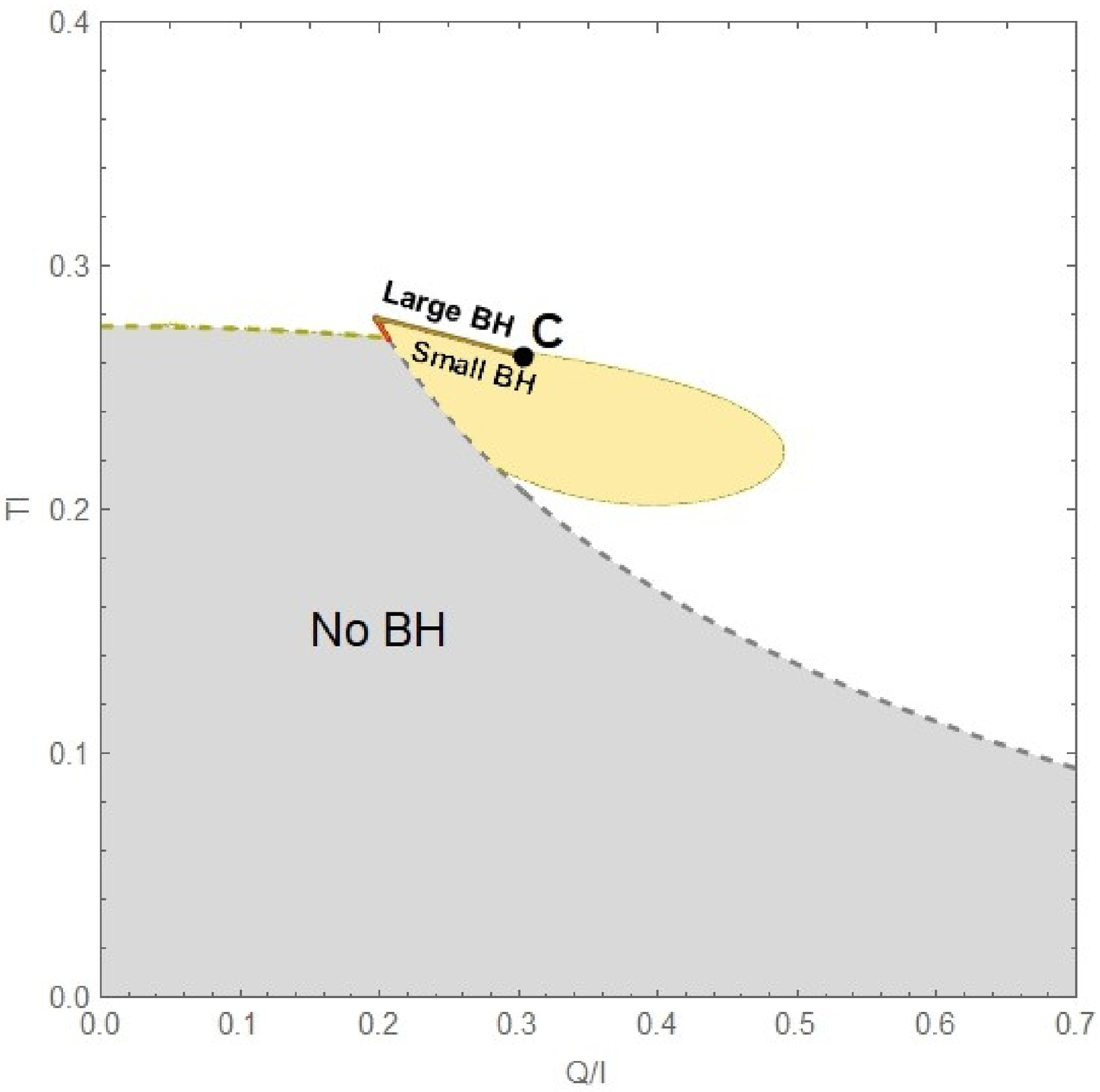}\label{fig:iDBIQT:d}}
\end{center}
\caption{{\footnotesize The phase diagrams in the $\tilde{Q}$-$\tilde{T}$
plane for iBI-AdS black holes with $a/l^{2}=0.01$ in $a/l^{2}=0.1$. The first
order phase transition lines separating large BH and small BH are displayed by
the brown lines, and they terminate at the critical points, marked by black
dots. There are also zeroth order phase transition lines, depicted by the red
lines. All phases in the diagram are thermally stable. However, the phases in
the yellow region are electrically unstable.}}%
\label{fig:iDBIQT}%
\end{figure}

The phase diagrams of iBI-AdS black holes for $a/l^{2}=0.01$ and $a/l^{2}%
=0.1$\ are displayed in the $\tilde{Q}$-$\tilde{T}$ plane in FIG.
\ref{fig:iDBIQT}, where we have the LBH/SBH first order phase transition lines
(the brown lines), the LBH/SBH zeroth order phase transition lines (the red
lines) and the critical points (the black dots). For $a/l^{2}=0.01$, FIGs.
\ref{fig:DBIa001} and \ref{fig:iDBIQT:a} shows that the phase diagram of
iBI-AdS black holes is similar to that of BI-AdS black holes. Moreover, FIG.
\ref{fig:iDBIQT:d} shows that the phase diagram of iBI-AdS black holes with
$a/l^{2}=0.1$ is similar to that with $a/l^{2}=0.01$, in the way that they
have the LBH/SBH first order and the zeroth order phase transitions. This is
expected from FIG. \ref{fig:iDBIc:a}, which shows that the $a/l^{2}=0.1$ line
intersects the critical line and the boundary $\tilde{Q}_{45}\left(  \tilde
{a}\right)  $. However, as shown in FIG. \ref{fig:DBIa010}, there no phase
transitions occurring in the phase diagram of BI-AdS black holes with
$a/l^{2}=0.1$.

All phases in FIG. \ref{fig:iDBIQT} are thermally stable. The black holes in
the yellow region are electrically unstable. As with the critical points of
BI-AdS black holes, the critical points of iBI-AdS black holes are also
electrically unstable, which are highlighted in FIG. \ref{fig:iDBIQT:c}. For
BI-AdS black holes, FIGs. \ref{fig:DBIa001} and \ref{fig:DBIa010} shows that
the black holes very close to the boundaries of NO BH regions are always
electrically unstable. However for iBI-AdS black holes, FIG. \ref{fig:iDBIQT}
shows that the black holes very close to the boundaries of NO BH regions can
be electrically stable for large enough values of $Q/l$.

\section{Conclusion}

\label{Sec:Con}

We have investigated the thermodynamic behavior of NLED AdS black holes in an
extended phase space, which includes the conjugate pressure/volume quantities,
any dimensionful couplings $a_{i}$ in NLED and their associated conjugates
$\mathcal{A}_{i}$. For a generic NLED black hole, we first computed its
Euclidean action to obtain the Gibbs free energy. To obtain consistency of the
Smarr relation, we found that it is necessary to include the conjugate pairs
$\left(  a_{i},\mathcal{A}_{i}\right)  $. It showed that the black hole's
temperature $T$, charge $Q$, horizon radius $r_{+}$ (thermodynamic volume
$V$), the AdS radius $l$ (pressure $P$) and the dimensionful couplings $a_{i}$
could be connected by
\begin{equation}
Tl=\tilde{T}\left(  r_{+}/l,Q/l,a_{i}l^{-c_{i}}\right)  ,
\end{equation}
where $c_{i}$ is the dimension of $a_{i}$. In the canonical ensemble with
fixed $T$ and $Q$, we found that the critical behavior and phase structure of
the black hole are determined by $\tilde{Q}\equiv Q/l$ and $\tilde{a}%
_{i}\equiv a_{i}l^{-c_{i}}$.

For BI-AdS black holes, we examined their critical behavior and phase
structure, whose dependence on $\tilde{Q}$ and $\tilde{a}$ was plotted in FIG.
\ref{fig:DBIQa}. There are six regions in FIG. \ref{fig:DBIQa}, and each
region has a different phase behavior. Specially, the LBH/SBH/LBH reentrant
phase transition occurs in Region IV. For iBI-AdS black holes, we displayed
the dependence of their critical behavior and phase structure on $\tilde{Q}$
and $\tilde{a}$ in FIG. \ref{fig:iDBIQa}, where there are also six regions,
and the LBH/SBH/LBH reentrant phase transition occurs in Region V. We
summarize the results of the critical behavior and phase structure for BI-AdS
and iBI-AdS black holes in Table \ref{tab:2}. \begin{table}[tbh]
\centering$%
\begin{tabular}
[c]{|m{1.5in}<{\centering}|m{2.3in}<{\centering}|m{2.3in}<{\centering}|}\hline
& {\footnotesize BI-AdS BH} & {\footnotesize iBI-AdS BH}\\\hline
{\footnotesize Critical line} & {\footnotesize The critical line has a
physical branch and an unphysical one, which have finite length and both
terminate at $\left\{  \tilde{a}_{c},\tilde{Q}_{c}\right\}  \simeq\left\{
0.069,0.37\right\}  $.} & {\footnotesize The critical line is a semi-infinite
line and extends to the infinity $\tilde{Q}=+\infty$.}\\\hline
{\footnotesize Reentrant phase transition region} & {\footnotesize This region
has a finite area and terminates at $\left\{  \tilde{a}_{c},\tilde{Q}%
_{c}\right\}  $.} & {\footnotesize This region extends to the infinity
$\tilde{Q}=+\infty$.}\\\hline
{\footnotesize Varying $P$ with fixed $Q$ and $a$ case ($\tilde{Q}_{l}\left(
\tilde{a}\right)  $ line)} & {\footnotesize There exists one physical critical
point for $Q/\sqrt{a}>\sqrt{2}$. The reentrant phase transition occurs for
$\sqrt{2}<Q/\sqrt{a}<2$.} & {\footnotesize There always exists one physical
critical point. The reentrant phase transition occurs for $Q/\sqrt{a}<2$%
.}\\\hline
{\footnotesize Varying $Q$ with fixed $P$ and $a$ case (constant-$\tilde{a}$
line)} & {\footnotesize A physical critical point and reentrant phase
transition occur for $\tilde{a}<\tilde{a}_{c}$.} & {\footnotesize A physical
critical point and reentrant phase transition occur for all values of $l$ and
$a$.}\\\hline
\end{tabular}
\ $\caption{{\small Critical behavior and phase structure for BI-AdS and
iBI-AdS black holes.}}%
\label{tab:2}%
\end{table}

The thermodynamically preferred phases, along with the zeroth and first phase
transitions and critical points, were displayed in FIGs. \ref{fig:DBIa001} and
\ref{fig:DBIa010} for BI-AdS black holes and in FIG. \ref{fig:iDBIQT} for
iBI-AdS black holes. We examined thermal and electrical stabilities of the
black holes and found that all the thermodynamically preferred phases are
thermal stable. However, the thermodynamically preferred phases in yellow
regions in these figures were found to be electrical unstable. The possible
equilibrium phases residing in the yellow regions were discussed in
\cite{IN-Chamblin:1999hg}, which listed extremal black holes, anti--de Sitter
space and black holes surrounded by a gas of particles as candidates. However,
this question still remains open. In
\cite{IN-Baggioli:2016oju,IN-Wang:2018hwg,Con-Cremonini:2017qwq}, the
electrical transport behavior of the dual theory has been discussed for BI-AdS
and iBI-AdS black holes in the context of gauge/gravity duality. It might be
inspiring to explore the possible equilibrium phases residing in yellow
regions from a holographic perspective.

\begin{acknowledgments}
We are grateful to Zheng Sun and Zhipeng Zhang for useful discussions and
valuable comments. This work is supported in part by NSFC (Grant No. 11005016,
11175039 and 11375121).
\end{acknowledgments}

\appendix

\section{Derivation of Smarr Relation}

In this appendix, we directly derive the Smarr relation from the definitions
of the thermodynamic quantities of the black hole. The Lagrangian
$\mathcal{L}\left(  s,a_{i}\right)  $ is a function of $s$ and the parameters
$a_{i}$. Performing the dimensional analysis, we find%
\begin{equation}
\left[  \mathcal{L}\right]  =L^{-2}\text{, }\left[  s\right]  =L^{-2}\text{,
}\left[  a_{i}\right]  =L^{c_{i}}\text{.}%
\end{equation}
Euler's theorem says%
\begin{equation}
2\mathcal{L}\left(  s,a_{i}\right)  =2s\mathcal{L}^{\prime}\left(
s,a_{i}\right)  -c_{i}a_{i}\frac{\partial\mathcal{L}\left(  s,a_{i}\right)
}{\partial a_{i}}.
\end{equation}
For a dimensionful coupling $a_{i}$, we have%
\begin{align}
c_{i}a_{i}\mathcal{A}_{i}  &  =-4\pi c_{i}a_{i}\frac{\partial}{\partial a_{i}%
}\int_{r_{+}}^{\infty}drr^{2}\mathcal{L}\left(  \frac{A_{t}^{\prime2}\left(
r\right)  }{2},a_{i}\right)  +c_{i}a_{i}Q\frac{\partial\Phi}{\partial a_{i}%
}\nonumber\\
&  =-c_{i}a_{i}\left[  Q\frac{\partial\Phi}{\partial a_{i}}+4\pi\int_{r_{+}%
}^{\infty}drr^{2}\frac{\partial\mathcal{L}\left(  \frac{A_{t}^{\prime2}\left(
r\right)  }{2},a_{i}\right)  }{\partial a_{i}}\right]  +c_{i}a_{i}%
Q\frac{\partial\Phi}{\partial a_{i}}\nonumber\\
&  =8\pi\int_{r_{+}}^{\infty}drr^{2}\left[  \mathcal{L}\left(  \frac
{A_{t}^{\prime2}\left(  r\right)  }{2},a_{i}\right)  -\frac{A_{t}^{\prime
2}\left(  r\right)  }{2}\frac{\partial\mathcal{L}\left(  \frac{A_{t}^{\prime
2}\left(  r\right)  }{2},a_{i}\right)  }{\partial s}\right] \\
&  =8\pi\int_{r_{+}}^{\infty}drr^{2}\mathcal{L}\left(  \frac{A_{t}^{\prime
2}\left(  r\right)  }{2},a_{i}\right)  -Q\Phi.\nonumber
\end{align}
Therefore, one has%
\begin{align*}
&  2\left(  TS-VP\right)  +c_{i}a_{i}\mathcal{A}_{i}+Q\Phi\\
&  =8\pi r_{+}\left\{  1+\frac{r_{+}^{2}}{l^{2}}+\frac{r_{+}^{2}}{2}\left[
\mathcal{L}\left(  \frac{A_{t}^{\prime2}\left(  r_{+}\right)  }{2}%
,a_{i}\right)  -A_{t}^{\prime}\left(  r_{+}\right)  \frac{Q}{r_{+}^{2}%
}\right]  \right\}  +8\pi\int_{r_{+}}^{\infty}drr^{2}\mathcal{L}\left(
s,a_{i}\right) \\
&  =M-Q\Phi+8\pi\left\{  \frac{3}{2}\int_{r_{+}}^{\infty}drr^{2}%
\mathcal{L}\left(  \frac{A_{t}^{\prime2}\left(  r\right)  }{2},a_{i}\right)
+\frac{r_{+}^{3}}{2}\left[  \mathcal{L}\left(  \frac{A_{t}^{\prime2}\left(
r_{+}\right)  }{2},a_{i}\right)  -A_{t}^{\prime}\left(  r_{+}\right)
\frac{Q\Phi}{r_{+}^{2}}\right]  \right\} \\
&  =M,
\end{align*}
where in the last equation, we use%
\begin{align*}
\int_{r_{+}}^{\infty}drr^{2}\mathcal{L}\left(  \frac{A_{t}^{\prime2}\left(
r\right)  }{2},a_{i}\right)   &  =-\frac{r_{+}^{3}\mathcal{L}\left(
\frac{A_{t}^{\prime2}\left(  r_{+}\right)  }{2},a_{i}\right)  }{3}-\frac{1}%
{3}\int_{r_{+}}^{\infty}drr^{3}\mathcal{L}^{\prime}\left(  \frac{A_{t}%
^{\prime2}\left(  r\right)  }{2},a_{i}\right)  A_{t}^{\prime}\left(  r\right)
A_{t}^{\prime\prime}\left(  r\right) \\
&  =-\frac{r_{+}^{3}\mathcal{L}\left(  s\right)  }{3}+\frac{q}{3}A_{t}%
^{\prime}\left(  r_{+}\right)  r_{+}+\frac{q\Phi}{12\pi}\text{.}%
\end{align*}


\begin{thebibliography}{99}                                                                                               %


\bibitem {IN-Hawking:1974sw}S.~W.~Hawking, ``Particle Creation by Black
Holes,'' Commun.\ Math.\ Phys.\ \textbf{43}, 199 (1975) Erratum:
[Commun.\ Math.\ Phys.\ \textbf{46}, 206 (1976)]. doi:10.1007/BF02345020, 10.1007/BF01608497

\bibitem {IN-Bekenstein:1972tm}J.~D.~Bekenstein, ``Black holes and the second
law,'' Lett.\ Nuovo Cim.\ \textbf{4}, 737 (1972). doi:10.1007/BF02757029

\bibitem {IN-Bekenstein:1973ur}J.~D.~Bekenstein, ``Black holes and entropy,''
Phys.\ Rev.\ D \textbf{7}, 2333 (1973). doi:10.1103/PhysRevD.7.2333

\bibitem {IN-Bardeen:1973gs}J.~M.~Bardeen, B.~Carter and S.~W.~Hawking, ``The
Four laws of black hole mechanics,'' Commun.\ Math.\ Phys.\ \textbf{31}, 161
(1973). doi:10.1007/BF01645742

\bibitem {IN-Maldacena:1997re}J.~M.~Maldacena, ``The Large N limit of
superconformal field theories and supergravity,''
Int.\ J.\ Theor.\ Phys.\ \textbf{38}, 1113 (1999)
[Adv.\ Theor.\ Math.\ Phys.\ \textbf{2}, 231 (1998)]
doi:10.1023/A:1026654312961, 10.4310/ATMP.1998.v2.n2.a1 [hep-th/9711200].

\bibitem {IN-Hawking:1982dh}S.~W.~Hawking and D.~N.~Page, ``Thermodynamics of
Black Holes in anti-De Sitter Space,'' Commun.\ Math.\ Phys.\ \textbf{87}, 577
(1983). doi:10.1007/BF01208266

\bibitem {IN-Witten:1998zw}E.~Witten, ``Anti-de Sitter space, thermal phase
transition, and confinement in gauge theories,''
Adv.\ Theor.\ Math.\ Phys.\ \textbf{2}, 505 (1998)
doi:10.4310/ATMP.1998.v2.n3.a3 [hep-th/9803131].

\bibitem {IN-Chamblin:1999tk}A.~Chamblin, R.~Emparan, C.~V.~Johnson and
R.~C.~Myers, \textquotedblleft Charged AdS black holes and catastrophic
holography,\textquotedblright\ Phys.\ Rev.\ D \textbf{60}, 064018 (1999)
doi:10.1103/PhysRevD.60.064018 [hep-th/9902170].

\bibitem {IN-Chamblin:1999hg}A.~Chamblin, R.~Emparan, C.~V.~Johnson and
R.~C.~Myers, \textquotedblleft Holography, thermodynamics and fluctuations of
charged AdS black holes,\textquotedblright\ Phys.\ Rev.\ D \textbf{60}, 104026
(1999) doi:10.1103/PhysRevD.60.104026 [hep-th/9904197].

\bibitem {IN-Dolan:2011xt}B.~P.~Dolan, ``Pressure and volume in the first law
of black hole thermodynamics,'' Class.\ Quant.\ Grav.\ \textbf{28}, 235017
(2011) doi:10.1088/0264-9381/28/23/235017 [arXiv:1106.6260 [gr-qc]].

\bibitem {IN-Kubiznak:2012wp}D.~Kubiznak and R.~B.~Mann, ``P-V criticality of
charged AdS black holes,'' JHEP \textbf{1207}, 033 (2012)
doi:10.1007/JHEP07(2012)033 [arXiv:1205.0559 [hep-th]].

\bibitem {IN-Kastor:2009wy}D.~Kastor, S.~Ray and J.~Traschen, ``Enthalpy and
the Mechanics of AdS Black Holes,'' Class.\ Quant.\ Grav.\ \textbf{26}, 195011
(2009) doi:10.1088/0264-9381/26/19/195011 [arXiv:0904.2765 [hep-th]].

\bibitem {IN-Wei:2012ui}S.~W.~Wei and Y.~X.~Liu, ``Critical phenomena and
thermodynamic geometry of charged Gauss-Bonnet AdS black holes,''
Phys.\ Rev.\ D \textbf{87}, no. 4, 044014 (2013)
doi:10.1103/PhysRevD.87.044014 [arXiv:1209.1707 [gr-qc]].

\bibitem {IN-Cai:2013qga}R.~G.~Cai, L.~M.~Cao, L.~Li and R.~Q.~Yang, ``P-V
criticality in the extended phase space of Gauss-Bonnet black holes in AdS
space,'' JHEP \textbf{1309}, 005 (2013) doi:10.1007/JHEP09(2013)005
[arXiv:1306.6233 [gr-qc]].

\bibitem {IN-Xu:2014kwa}W.~Xu and L.~Zhao, ``Critical phenomena of static
charged AdS black holes in conformal gravity,'' Phys.\ Lett.\ B \textbf{736},
214 (2014) doi:10.1016/j.physletb.2014.07.019 [arXiv:1405.7665 [gr-qc]].

\bibitem {IN-Frassino:2014pha}A.~M.~Frassino, D.~Kubiznak, R.~B.~Mann and
F.~Simovic, ``Multiple Reentrant Phase Transitions and Triple Points in
Lovelock Thermodynamics,'' JHEP \textbf{1409}, 080 (2014)
doi:10.1007/JHEP09(2014)080 [arXiv:1406.7015 [hep-th]].

\bibitem {IN-Dehghani:2014caa}M.~H.~Dehghani, S.~Kamrani and A.~Sheykhi,
``$P-V$ criticality of charged dilatonic black holes,'' Phys.\ Rev.\ D
\textbf{90}, no. 10, 104020 (2014) doi:10.1103/PhysRevD.90.104020
[arXiv:1505.02386 [hep-th]].

\bibitem {IN-Hennigar:2015esa}R.~A.~Hennigar, W.~G.~Brenna and R.~B.~Mann,
``$P ? v$ criticality in quasitopological gravity,'' JHEP \textbf{1507}, 077
(2015) doi:10.1007/JHEP07(2015)077 [arXiv:1505.05517 [hep-th]].

\bibitem {IN-Gunasekaran:2012dq}S.~Gunasekaran, R.~B.~Mann and D.~Kubiznak,
``Extended phase space thermodynamics for charged and rotating black holes and
Born-Infeld vacuum polarization,'' JHEP \textbf{1211}, 110 (2012)
doi:10.1007/JHEP11(2012)110 [arXiv:1208.6251 [hep-th]].

\bibitem {IN-Altamirano:2013ane}N.~Altamirano, D.~Kubiznak and R.~B.~Mann,
``Reentrant phase transitions in rotating anti--de Sitter black holes,''
Phys.\ Rev.\ D \textbf{88}, no. 10, 101502 (2013)
doi:10.1103/PhysRevD.88.101502 [arXiv:1306.5756 [hep-th]].

\bibitem {IN-Zou:2016sab}D.~C.~Zou, R.~Yue and M.~Zhang, ``Reentrant phase
transitions of higher-dimensional AdS black holes in dRGT massive gravity,''
Eur.\ Phys.\ J.\ C \textbf{77}, no. 4, 256 (2017)
doi:10.1140/epjc/s10052-017-4822-9 [arXiv:1612.08056 [gr-qc]].

\bibitem {IN-Hennigar:2015wxa}R.~A.~Hennigar and R.~B.~Mann, ``Reentrant phase
transitions and van der Waals behaviour for hairy black holes,'' Entropy
\textbf{17}, no. 12, 8056 (2015) doi:10.3390/e17127862 [arXiv:1509.06798 [hep-th]].

\bibitem {IN-Soleng:1995kn}H.~H.~Soleng, ``Charged black points in general
relativity coupled to the logarithmic U(1) gauge theory,'' Phys.\ Rev.\ D
\textbf{52}, 6178 (1995) doi:10.1103/PhysRevD.52.6178 [hep-th/9509033].

\bibitem {IN-AyonBeato:1998ub}E.~Ayon-Beato and A.~Garcia, \textquotedblleft
Regular black hole in general relativity coupled to nonlinear
electrodynamics,\textquotedblright\ Phys.\ Rev.\ Lett.\ \textbf{80}, 5056
(1998) doi:10.1103/PhysRevLett.80.5056 [gr-qc/9911046].

\bibitem {IN-Maeda:2008ha}H.~Maeda, M.~Hassaine and C.~Martinez, ``Lovelock
black holes with a nonlinear Maxwell field,'' Phys.\ Rev.\ D \textbf{79},
044012 (2009) doi:10.1103/PhysRevD.79.044012 [arXiv:0812.2038 [gr-qc]].

\bibitem {IN-Hendi:2017mgb}S.~H.~Hendi, B.~Eslam Panah, S.~Panahiyan and
A.~Sheykhi, \textquotedblleft Dilatonic BTZ black holes with power-law
field,\textquotedblright\ Phys.\ Lett.\ B \textbf{767}, 214 (2017)
doi:10.1016/j.physletb.2017.01.066 [arXiv:1703.03403 [gr-qc]].

\bibitem {IN-Tao:2017fsy}J.~Tao, P.~Wang and H.~Yang, ``Testing holographic
conjectures of complexity with Born--Infeld black holes,'' Eur.\ Phys.\ J.\ C
\textbf{77}, no. 12, 817 (2017) doi:10.1140/epjc/s10052-017-5395-3
[arXiv:1703.06297 [hep-th]].

\bibitem {IN-Guo:2017bru}X.~Guo, P.~Wang and H.~Yang, ``Membrane Paradigm and
Holographic DC Conductivity for Nonlinear Electrodynamics,'' Phys.\ Rev.\ D
\textbf{98}, no. 2, 026021 (2018) doi:10.1103/PhysRevD.98.026021
[arXiv:1711.03298 [hep-th]].

\bibitem {IN-Mu:2017usw}B.~Mu, P.~Wang and H.~Yang, ``Holographic DC
Conductivity for a Power-law Maxwell Field,'' arXiv:1711.06569 [hep-th].

\bibitem {IN-Hendi:2012um}S.~H.~Hendi and M.~H.~Vahidinia, ``Extended phase
space thermodynamics and P-V criticality of black holes with a nonlinear
source,'' Phys.\ Rev.\ D \textbf{88}, no. 8, 084045 (2013)
doi:10.1103/PhysRevD.88.084045 [arXiv:1212.6128 [hep-th]].

\bibitem {IN-Mo:2016jqd}J.~X.~Mo, G.~Q.~Li and X.~B.~Xu, ``Effects of
power-law Maxwell field on the critical phenomena of higher dimensional
dilaton black holes,'' Phys.\ Rev.\ D \textbf{93}, no. 8, 084041 (2016)
doi:10.1103/PhysRevD.93.084041 [arXiv:1601.05500 [gr-qc]].

\bibitem {IN-Yerra:2018mni}P.~K.~Yerra and C.~Bhamidipati, ``A Note on
Critical Nonlinearly Charged Black Holes,'' arXiv:1806.08226 [hep-th].

\bibitem {IN-Nam:2018tpf}C.~H.~Nam, ``Non-linear charged dS black hole and its
thermodynamics and phase transitions,'' Eur.\ Phys.\ J.\ C \textbf{78}, no. 5,
418 (2018). doi:10.1140/epjc/s10052-018-5922-x

\bibitem {IN-Dey:2004yt}T.~K.~Dey, \textquotedblleft Born-Infeld black holes
in the presence of a cosmological constant,\textquotedblright\ Phys.\ Lett.\ B
\textbf{595}, 484 (2004) doi:10.1016/j.physletb.2004.06.047 [hep-th/0406169].

\bibitem {IN-Cai:2004eh}R.~G.~Cai, D.~W.~Pang and A.~Wang, \textquotedblleft
Born-Infeld black holes in (A)dS spaces,\textquotedblright\ Phys.\ Rev.\ D
\textbf{70}, 124034 (2004) doi:10.1103/PhysRevD.70.124034 [hep-th/0410158].

\bibitem {IN-Fernando:2003tz}S.~Fernando and D.~Krug, ``Charged black hole
solutions in Einstein-Born-Infeld gravity with a cosmological constant,''
Gen.\ Rel.\ Grav.\ \textbf{35}, 129 (2003) doi:10.1023/A:1021315214180 [hep-th/0306120].

\bibitem {IN-Fernando:2006gh}S.~Fernando, ``Thermodynamics of
Born-Infeld-anti-de Sitter black holes in the grand canonical ensemble,''
Phys.\ Rev.\ D \textbf{74}, 104032 (2006) doi:10.1103/PhysRevD.74.104032 [hep-th/0608040].

\bibitem {IN-Banerjee:2010da}R.~Banerjee, S.~Ghosh and D.~Roychowdhury, ``New
type of phase transition in Reissner Nordstrom--AdS black hole and its
thermodynamic geometry,'' Phys.\ Lett.\ B \textbf{696}, 156 (2011)
doi:10.1016/j.physletb.2010.12.010 [arXiv:1008.2644 [gr-qc]].

\bibitem {IN-Banerjee:2011cz}R.~Banerjee and D.~Roychowdhury, ``Critical
phenomena in Born-Infeld AdS black holes,'' Phys.\ Rev.\ D \textbf{85}, 044040
(2012) doi:10.1103/PhysRevD.85.044040 [arXiv:1111.0147 [gr-qc]].

\bibitem {IN-Lala:2011np}A.~Lala and D.~Roychowdhury, ``Ehrenfest's scheme and
thermodynamic geometry in Born-Infeld AdS black holes,'' Phys.\ Rev.\ D
\textbf{86}, 084027 (2012) doi:10.1103/PhysRevD.86.084027 [arXiv:1111.5991 [gr-qc]].

\bibitem {IN-Banerjee:2012zm}R.~Banerjee and D.~Roychowdhury, ``Critical
behavior of Born Infeld AdS black holes in higher dimensions,'' Phys.\ Rev.\ D
\textbf{85}, 104043 (2012) doi:10.1103/PhysRevD.85.104043 [arXiv:1203.0118 [gr-qc]].

\bibitem {IN-Azreg-Ainou:2014twa} M.~Azreg-Ainou,  ``Black hole
thermodynamics: No inconsistency via the inclusion of the missing $P-V$
terms,''  Phys.\ Rev.\ D \textbf{91}, 064049 (2015)
doi:10.1103/PhysRevD.91.064049  [arXiv:1411.2386 [gr-qc]].

\bibitem {IN-Hendi:2015hoa}S.~H.~Hendi, B.~Eslam Panah and S.~Panahiyan,
``Einstein-Born-Infeld-Massive Gravity: adS-Black Hole Solutions and their
Thermodynamical properties,'' JHEP \textbf{1511}, 157 (2015)
doi:10.1007/JHEP11(2015)157 [arXiv:1508.01311 [hep-th]].

\bibitem {IN-Zangeneh:2016fhy}M.~Kord Zangeneh, A.~Dehyadegari,
M.~R.~Mehdizadeh, B.~Wang and A.~Sheykhi, ``Thermodynamics, phase transitions
and Ruppeiner geometry for Einstein--dilaton--Lifshitz black holes in the
presence of Maxwell and Born--Infeld electrodynamics,'' Eur.\ Phys.\ J.\ C
\textbf{77}, no. 6, 423 (2017) doi:10.1140/epjc/s10052-017-4989-0
[arXiv:1610.06352 [hep-th]].

\bibitem {IN-Zeng:2016sei}X.~X.~Zeng, X.~M.~Liu and L.~F.~Li, ``Phase
structure of the Born--Infeld--anti-de Sitter black holes probed by non-local
observables,'' Eur.\ Phys.\ J.\ C \textbf{76}, no. 11, 616 (2016)
doi:10.1140/epjc/s10052-016-4463-4 [arXiv:1601.01160 [hep-th]].

\bibitem {IN-Li:2016nll}S.~Li, H.~Lu and H.~Wei, ``Dyonic (A)dS Black Holes in
Einstein-Born-Infeld Theory in Diverse Dimensions,'' JHEP \textbf{1607}, 004
(2016) doi:10.1007/JHEP07(2016)004 [arXiv:1606.02733 [hep-th]].

\bibitem {IN-Zou:2013owa}D.~C.~Zou, S.~J.~Zhang and B.~Wang, \textquotedblleft
Critical behavior of Born-Infeld AdS black holes in the extended phase space
thermodynamics,\textquotedblright\ Phys.\ Rev.\ D \textbf{89}, no. 4, 044002
(2014) doi:10.1103/PhysRevD.89.044002 [arXiv:1311.7299 [hep-th]].

\bibitem {IN-Dehyadegari:2017hvd}A.~Dehyadegari and A.~Sheykhi,
\textquotedblleft Reentrant phase transition of Born-Infeld-AdS black
holes,\textquotedblright\ Phys.\ Rev.\ D \textbf{98}, no. 2, 024011 (2018)
doi:10.1103/PhysRevD.98.024011 [arXiv:1711.01151 [gr-qc]].

\bibitem {IN-Baggioli:2016oju}M.~Baggioli and O.~Pujolas, \textquotedblleft On
Effective Holographic Mott Insulators,\textquotedblright\ JHEP \textbf{1612},
107 (2016) doi:10.1007/JHEP12(2016)107 [arXiv:1604.08915 [hep-th]].

\bibitem {IN-Wang:2018hwg}P.~Wang, H.~Wu and H.~Yang, \textquotedblleft
Holographic DC Conductivity for Backreacted Nonlinear Electrodynamics with
Momentum Dissipation,\textquotedblright\ arXiv:1805.07913 [hep-th].

\bibitem {NELDBH-Hawking:1978jz}S.~W.~Hawking, ``Quantum Gravity and Path
Integrals,'' Phys.\ Rev.\ D \textbf{18}, 1747 (1978). doi:10.1103/PhysRevD.18.1747

\bibitem {NELDBH-Gibbons:2002du}G.~Gibbons, ``Euclidean quantum gravity: The
view from 2002,'' in The Future of Theoretical Physics and Cosmology.

\bibitem {NELDBH-Chamblin:1999hg}A.~Chamblin, R.~Emparan, C.~V.~Johnson and
R.~C.~Myers, ``Holography, thermodynamics and fluctuations of charged AdS
black holes,'' Phys.\ Rev.\ D \textbf{60}, 104026 (1999)
doi:10.1103/PhysRevD.60.104026 [hep-th/9904197].

\bibitem {NELDBH-Balasubramanian:1999re}V.~Balasubramanian and P.~Kraus, ``A
Stress tensor for Anti-de Sitter gravity,''
Commun.\ Math.\ Phys.\ \textbf{208}, 413 (1999) doi:10.1007/s002200050764 [hep-th/9902121].

\bibitem {NELDBH-Emparan:1999pm}R.~Emparan, C.~V.~Johnson and R.~C.~Myers,
``Surface terms as counterterms in the AdS / CFT correspondence,''
Phys.\ Rev.\ D \textbf{60}, 104001 (1999) doi:10.1103/PhysRevD.60.104001 [hep-th/9903238].

\bibitem {NELDBH-Olea:2005gb}R.~Olea, ``Mass, angular momentum and
thermodynamics in four-dimensional Kerr-AdS black holes,'' JHEP \textbf{0506},
023 (2005) doi:10.1088/1126-6708/2005/06/023 [hep-th/0504233].

\bibitem {NELDBH-Olea:2006vd}R.~Olea, ``Regularization of odd-dimensional AdS
gravity: Kounterterms,'' JHEP \textbf{0704}, 073 (2007)
doi:10.1088/1126-6708/2007/04/073 [hep-th/0610230].

\bibitem {NELDBH-Miskovic:2008ck}O.~Miskovic and R.~Olea, ``Thermodynamics of
Einstein-Born-Infeld black holes with negative cosmological constant,''
Phys.\ Rev.\ D \textbf{77}, 124048 (2008) doi:10.1103/PhysRevD.77.124048
[arXiv:0802.2081 [hep-th]].

\bibitem {NELDBH-Kastor:2010gq}D.~Kastor, S.~Ray and J.~Traschen, ``Smarr
Formula and an Extended First Law for Lovelock Gravity,''
Class.\ Quant.\ Grav.\ \textbf{27}, 235014 (2010)
doi:10.1088/0264-9381/27/23/235014 [arXiv:1005.5053 [hep-th]].

\bibitem {NELDBH-Kastor:2009wy}D.~Kastor, S.~Ray and J.~Traschen,
\textquotedblleft Enthalpy and the Mechanics of AdS Black
Holes,\textquotedblright\ Class.\ Quant.\ Grav.\ \textbf{26}, 195011 (2009)
doi:10.1088/0264-9381/26/19/195011 [arXiv:0904.2765 [hep-th]].

\bibitem {Con-Cremonini:2017qwq}S.~Cremonini, A.~Hoover and L.~Li,
\textquotedblleft Backreacted DBI Magnetotransport with Momentum
Dissipation,\textquotedblright\ JHEP \textbf{1710}, 133 (2017)
doi:10.1007/JHEP10(2017)133 [arXiv:1707.01505 [hep-th]].
\end{thebibliography}
\end{document}